\documentclass[aps,prl,groupedaddress,amsmath,amssymb,twocolumn,reprint]{revtex4-2}
\usepackage{physics}
\usepackage{graphicx}
\usepackage{threeparttable}

\begin{document}
\newcommand{\dif}{\mathrm{d}}
\title{On-demand storage and retrieval of microwave photons using a superconducting multi-resonator quantum memory} 

\author{Zenghui Bao}

\author{Zhiling Wang}

\author{Yukai Wu}

\author{Yan Li}

\author{Cheng Ma}

\author{Yipu Song}

\author{Hongyi Zhang}
\email{hyzhang2016@tsinghua.edu.cn}

\author{Luming Duan}
\email{lmduan@tsinghua.edu.cn}

\affiliation{Center for Quantum Information, Institute for Interdisciplinary Information Sciences, Tsinghua University, Beijing 100084, PR China}

\date{\today}

\begin{abstract}

A quantum memory that can store quantum states faithfully and retrieve them on demand has wide applications in quantum information science. An efficient quantum memory in the microwave regime working alongside quantum processors based on superconducting quantum circuits may serve as an important architecture for quantum computers. Here we realize on-demand storage and retrieval of weak coherent microwave photon pulses at the single-photon level. We implement a superconducting multi-resonator quantum memory which is composed of a set of frequency-tunable coplanar transmission line (CPW) resonators. By dynamically tuning the resonant frequencies of the resonators, we achieve tunable memory bandwidth from $10\,$MHz to $55\,$MHz, with an overall storage efficiency up to $12\,\%$ with well preserved phase coherence. We further demonstrate on-demand storage and retrieval of a time-bin flying qubit. This result opens up a prospect to integrate our chip-based quantum memory with the state-of-the-art superconducting quantum circuit technology for quantum information processing.

\end{abstract}

\maketitle 


Quantum memory is regarded as a key element of quantum communication and quantum computing systems~\cite{Kimble2008,Wolfgang2009}. Optical quantum memory based on either solid state system or laser-cooled ensemble of atoms can store multiple qubit states~\cite{Duan2017}, quantum entanglement~\cite{Gisin2011,Tittel2011}, coherent state and Fock state photons~\cite{Lvovsky2009,Sellars2010,Hugues2014,Guo2015,Tualle2019} with high fidelity and high efficiency~\cite{Guo2012,Zhu2019,Guo2020}, which enables long-distance quantum communication and the upcoming quantum network~\cite{Duan2001,Kimble2008,Hanson2018}.
Quantum memory in the microwave regime has also been broadly investigated aiming at realizing quantum interface between flying and stationary qubits or integrating quantum storage in quantum computing devices~\cite{Devoret2013,Kurizki2015}. A fully functional microwave quantum memory will be an integrated part of the von-Neumann architecture for quantum computing systems based on superconducting qubit processors~\cite{Kubo2011,Schuster2017}.

Spins in solid state systems have been considered for implementing a microwave quantum memory because of their long coherence time. It has been demonstrated that an ensemble of spins can be coherently coupled to microwave photon field \cite{Imamo2009,Zhu2011,Majer2011,Julsgaard2013,Jiang2015,Bertet2020}, hence realizing photon storage even down to the single photon level with a long storage time \cite{Kubo2011,Grezes2014,Grezes2015,Bertet2020}. However, it is challenging to improve the efficiency of the hybrid quantum interface \cite{Kurizki2015}, and typically one has a relatively low overall storage efficiency on the order of 0.1$\%$. Microwave resonators with high quality factors are a natural choice for the storage of microwave photons \cite{Wolfgang2009, Mallet2015, Schoelkopf2016}, with storage times in the millisecond range for three-dimensional (3D) superconducting resonators \cite{Schoelkopf2016}. A high efficiency interface between propagating photon field and the resonator mode can be realized by tailoring the coupling properties \cite{Martinis2013,Martinis2014}. In addition, microwave resonators can be strongly coupled to the qubits, like superconduting qubit and Rydberg atoms, which naturally forms an effective quantum interface between the photons and the qubit \cite{Schuster2017}. Moreover, carefully designed photon states in a microwave resonator can be directly used as a qubit, which is known to be robust to noise and useful in quantum computing and quantum metrology~\cite{Mirrahimi2014,Weigand2017}.

In order to realize broadband and multimode storage, a recent work proposes to use an array of  spatially and spectrally equally spaced resonators as a quantum memory to store propagating microwave photons \cite{Moiseev2017}.
The basic idea is akin to optical quantum memories based on photon echo and atomic frequency combs (AFC) \cite{Afzelius2009}. Photon field stored in different resonator modes with a certain phase difference will rephase into the commonly coupled continuous mode as an echo signal. An initial implementation of this scheme has been recently carried out with 3D resonators for classically strong microwave signals \cite{Moiseev2018}.

In this work, we report realization of a multi-resonator memory based on superconducting coplanar waveguide (CPW) structures, which is capable of on-demand storage and retrieval of microwave photons at the single photon level, with a storage efficiency up to $12\%$.
We also show that the device is capable of storing several input photon modes, and thus suitable to store a microwave time-bin flying qubit. One could therefore envision an on chip integration of our device with superconducting quantum processor to facilitate complex computing sequences.

\begin{figure}[!tbp]
\centering
\includegraphics[width=1\linewidth]{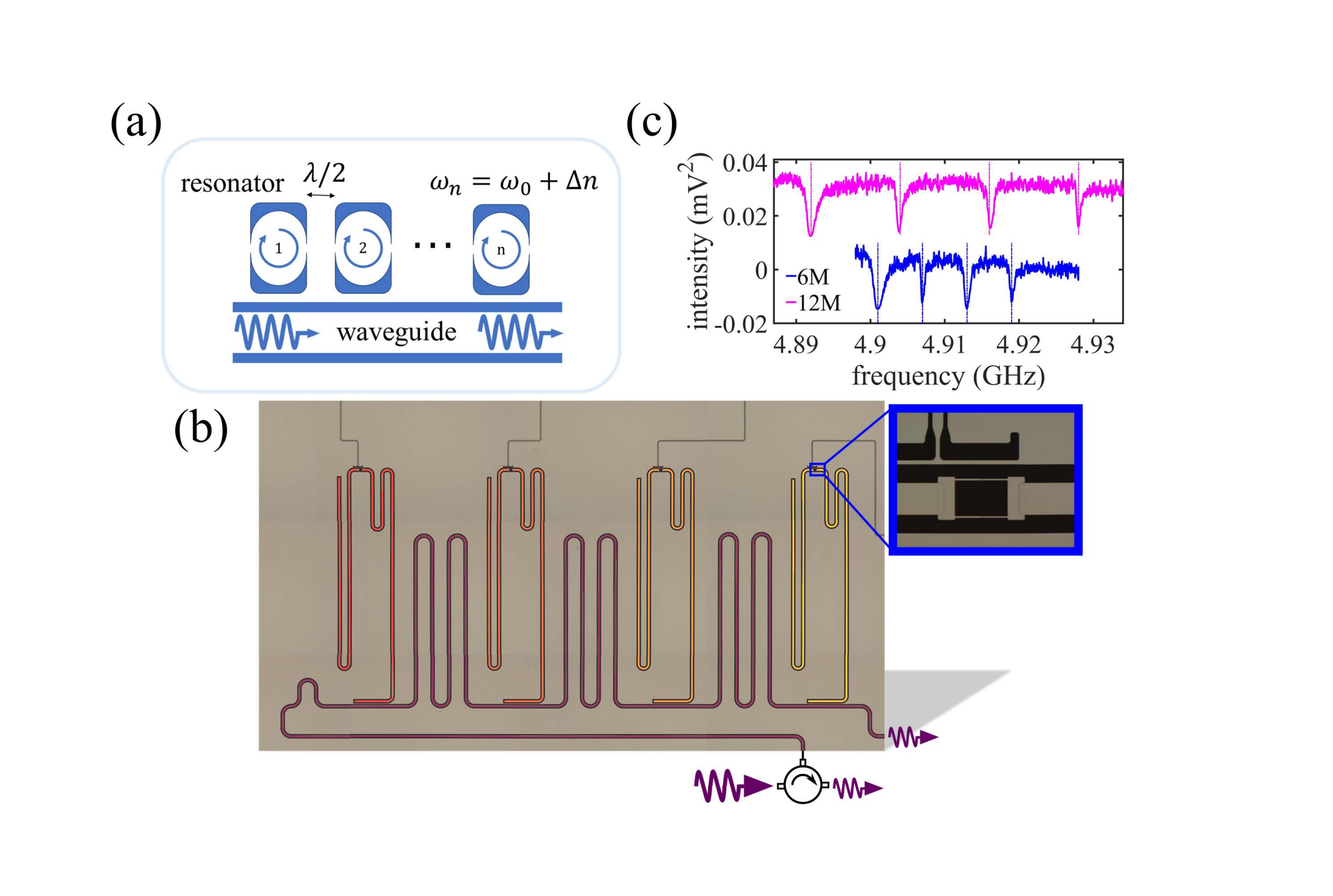}
\caption{\textbf{Sample illustration. }(a) Cartoon illustration of the multi resonator design. (b) Micrograph of the multiresonator sample used in the experiment. Our sample contains four resonators equally placed along the transmission line with a separation of $12.08\,$mm. (c) Transmission spectra of the resonators based frequency combs centered at $\omega_c = 4.91\,$GHz and with a periodicity $\Delta = 6\,$MHz or $12\,$MHz. The dashed lines mark the correspondingly ideal spectral positions.}
\label{samplespec}
\end{figure}

We consider a set of microwave resonators coupled to a common waveguide, as illustrated in Fig.~\ref{samplespec}(a). The resonators are periodically placed along the waveguide, with a spatial separation of $\lambda/2$. Their resonant frequencies are also equally spaced with a spectral separation of $\Delta$ and centered at $\omega_c = 2\pi c/\lambda$.
If a single photon field with a center frequency at $\omega_c$ and spectral distribution of $\delta_p$ is injected into the waveguide, the photon can be absorbed by the multiresonator as an collective excitation. Once being absorbed, it cannot immediately leak out from the resonators since the photon state in each resonator acquires a relative phase $e^{i\delta_n t}$, depending on the detuning $\delta_n = n\Delta$ of each resonator. The phase term leads to destructive interference, which makes the multiresonator system non-radiative until the rephasing condition is satisfied, corresponding to a time delay of $t = 1/\Delta$ \cite{Afzelius2009}. Therefore, such a multiresonator structure can be used as a microwave quantum memory. A detailed theoretical treatment can be found in Ref. \cite{Moiseev2017}.

Here, we use a device containing four superconducting coplanar waveguide (CPW) resonators coupled to a common transmission line, as illustrated in Fig.~\ref{samplespec}(b). The resonators are frequency tunable by embedding a SQUID in the center conductor of each resonator, giving rise to a typical tuning range of more than 200 MHz away from the sweet spot and a tuning speed better than $1\,$ns \cite{Delsing2008}.
The measured out-coupling rates of the resonators $\kappa_c/2\pi$ range from $0.25\,$MHz to $0.86\,$MHz, and the internal loss rates $\kappa_i/2\pi$ are on the order of $8\times10^4$ Hz at the sweet spot~\cite{supp}. 
With a careful calibration on the flux line of SQUID, the resonators can be spectrally positioned with a precision better than $100\,$kHz~\cite{supp}, hence a frequency comb can be readily realized.
Since there existing an effective coupling among the resonators, the resonator frequencies can not be precisely controlled when the periodicity of the frequency comb $\Delta$ is small~\cite{supp}.
In this experiments we have realized frequency combs containing four frequency components with the center frequency $\omega_c = 4.91\,$GHz and the periodicity $\Delta$ ranging from $3.5\,$MHz to $18\,$MHz. Some examples can be found in Fig.~\ref{samplespec}(c).

\begin{figure}[!tbp]
\centering
\includegraphics[width=1\linewidth]{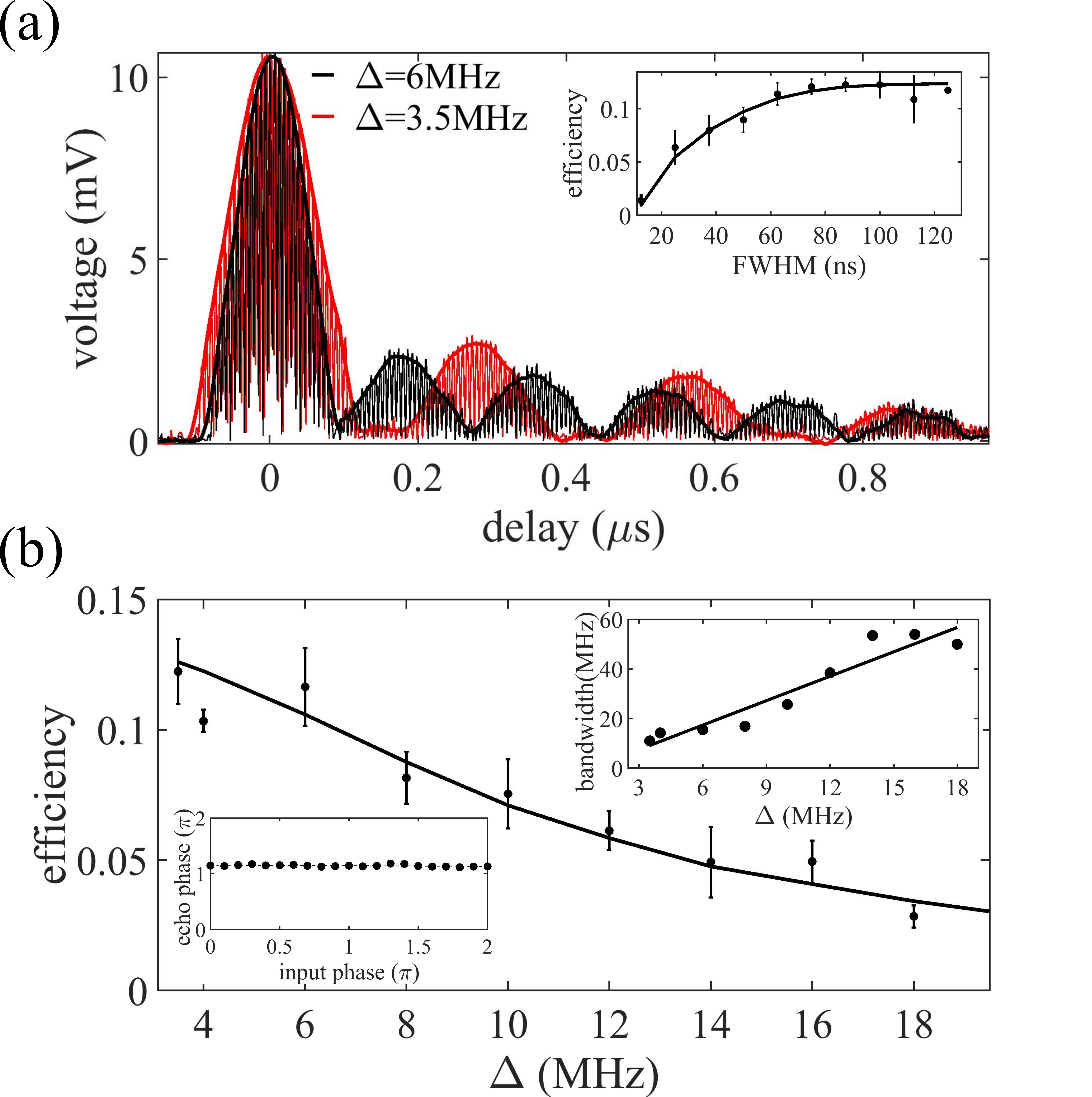}
\caption{\textbf{Photon echo from the multiresonators.} (a) Transmission signal from the multiresonator with $\Delta=3.5\,$MHz and $6\,$MHz, when sending a $4.91\,$GHz Gaussian enveloped pulse to the resonators. The input pulse width is $230\,$ns with FWHM of $97\,$ns for $\Delta=3.5\,$MHz, and $190\,$ns with FWHM of $78\,$ns for $\Delta=6\,$MHz. The output signal is down converted to $80\,$MHz and then acquired by an analog to digital converter. The oscillating thin lines are absolute value of the output voltage. The thick lines show envelop of the oscillating signal. The red and black lines represent transmission of a frequency comb with $\Delta = 3.5\,$MHz and $6\,$MHz, respectively. Inset of (a) shows storage efficiency as a function of FWHM of input pulse for $\Delta=3.5\,$MHz. The solid line shows numerical simulation result, which agrees well with the experiment. (b) The best achieved storage efficiencies for different $\Delta$. The solid line gives result from numerical simulation. The inverse of the smallest FWHM of input pulse to achieve the best achieved storage efficiency for varied $\Delta$ is defined as bandwidth, which is shown in upper right inset of (b). The solid line is a linear fit result. Lower left inset of (b) phase difference between the input pulse and the first echo pulse, with varied phase of input pulse. The dashed line is a linear fit. The error bars are given by a standard deviation of the corresponding measurement results.}
\label{echo}
\end{figure}

In order to characterize the photon echo process, we send a Gaussian-shaped coherent pulse into the transmission line, with an average photon number $ \langle n \rangle \sim 1$ and center frequency of $4.91\,$GHz. 
Fig.~\ref{echo}(a) shows transmission output for $\Delta = 3.5\,$MHz or $6\,$MHz. The first peak centered at $ t=0\,\mu$s is the directly transmitted signal that not being captured by the resonators. The following echo signal composes several peaks located at around $m/\Delta$, where $m$ is a positive integer indicating the order of echo signal.
Theoretically, an ideal photon release process with only one echo peak occurs when the impedance matching condition between the periodicity $\Delta$ and the out-coupling rate $\kappa_c$ is met, which is given by $\Delta =\pi\kappa_c/2$.
The impedance matching condition leads to an optimized interference effect among the resonators, which facilitates the best-achieved photon echo from the multiresonator, and thus the optimal storage efficiency~\cite{Moiseev2017,Moiseev2018}.
For our sample the impedance matching required $\Delta$ is about $1.5\,$MHz, and the appearance of higher order echo peaks in Fig.~\ref{echo}(a) is thus expected and also reproduced in our numerical simulation.

In order to evaluate storage efficiency and bandwidth of the memory, we use input pulses with varied full width at half maximum (FWHM) and measure overall storage efficiency. A typical result for $\Delta = 3.5\,$MHz is shown in the inset of  Fig.~\ref{echo}(a). Since transmission and reflection is symmetric for our multiresonator design, the overall storage efficiency is defined as the sum of the power of the first echo ($m=1$) in transmission and reflection signal divided by the strength of the input signal (see more details in Supplementary Information). The best-achieved storage efficiency is about $12\%$ when FWHM of the input pulse exceeds $100\,$ns, corresponding to a bandwidth of $10\,$MHz, which is close to the spectral range of the frequency comb $3\Delta$. A shorter pulse that exceeds the spectral range of the frequency comb can not be effectively captured by the multiresonator, leading to a degraded storage efficiency.
Such a feature is also confirmed in our numerical simulation~\cite{supp}.

We further measure the best-achieved storage efficiency and the corresponding bandwidth for different $\Delta$, and the result is shown in Fig.~\ref{echo}(b). With an increasing comb periodicity $\Delta$ from $3.5\,$MHz to $18\,$MHz, the storage bandwidth increases almost linearly from $10\,$MHz to about $55\,$MHz, with a sacrificed storage efficiency to about $3\%$ for $\Delta=18\,$MHz. It is worth noting that the best-achieved storage efficiency of our device can be further improved by using an optimized $\Delta$ given by the impedance matching condition. The effective coupling among the resonators hinders us to have a well aligned frequency comb with smaller $\Delta$, but would not infect the optimal storage efficiency, as discussed in Supplementary Information. An improved way to determine the intrinsic resonant frequencies of the resonators in the presence of the effective coupling would help to achieve higher storage efficiency.

The heterodyne detection scheme used in our experiment enables us to check phase coherence of the echo signal \cite{Martinis2013,Grezes2015,Bertet2020}. As shown in the lower left inset of Fig.~\ref{echo}(b), with varied phase from $0$ to $2\pi$ of input signal, the relative phase difference between the first echo and input signal keeps unchanged, which clearly demonstrates a good phase coherence of the storage and release process.

\begin{figure}[htbp]
\centering
\includegraphics[width=1\linewidth]{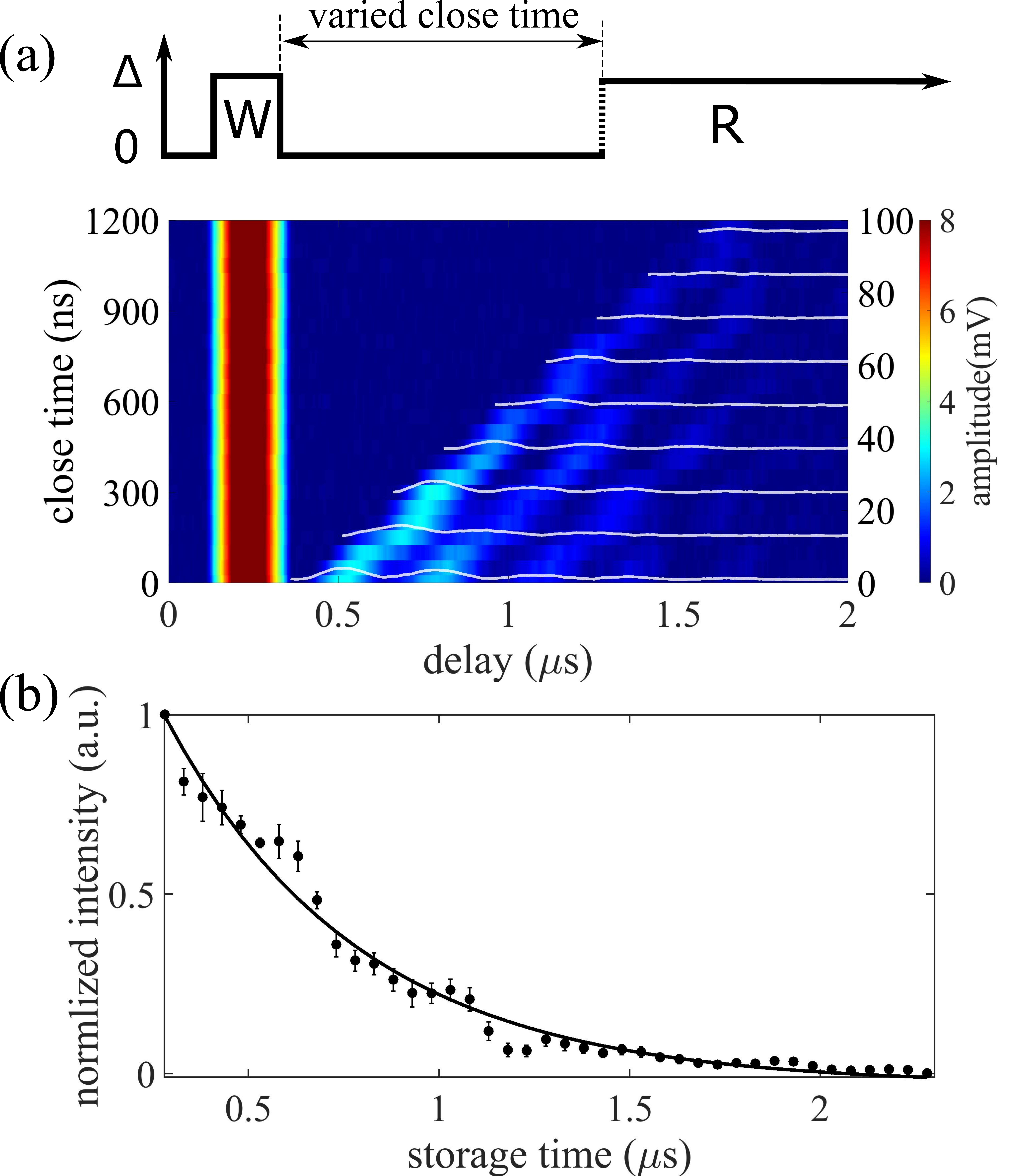}
\caption{\textbf{On-demand retrieval of the stored signal.} (a) An intensity map of the photon echo amplitude with varied close time. Crosscuts at the corresponding close times show echo intensity as a function of delay, which are plots with white solid lines. The input pulse width is $250\,$ns with FWHM of $110\,$ns. The top panel shows sequence of the resonator detuning setting with $\Delta=3.5\,$MHz. Either the input pulse capture or the echo signal release requires the multiresonator is set to a frequency comb with a periodicity of $\Delta$. The absorbed photons can be controllably stored in the multiresonator by tuning resonators into resonance. (b)The scattered plots are normalized echo intensity as a function of the storage time. The solid line is an exponential fitting result, which gives a time constant of about $0.51\,\mu$s. The error bars are given by a standard deviation of the measurement results.}
\label{on-demand}
\end{figure}

A prominent feature of our device is that the stored photons can be retrieved on-demand by dynamically tuning the frequencies of the resonators, and thus realizing a tunable storage time. 
By setting the resonant frequencies of the resonators to the same value, photon exchange between the multiresonator and the transmission line would be `freezed' and the photon state would keep in the resonators \cite{Moiseev2017}. In turn, if the frequencies of the resonators are restored to the original frequency comb, the stored photon field would continue to evolve and emit to the transmission line as a photon echo.

The flux sequence to realize on-demand retrieval is illustrated in the top panel of Fig.~\ref{on-demand}(a). `Writing' the input signal into the memory requires the multiresonator being set to a frequency comb with suitable $\Delta$. When the input signal has been absorbed by the multiresonator, the resonators are aligned to the same frequency to `close' the frequency comb and temporally stop the rephasing process among the photons in different resonators, where the stored photons form a non-radiative dark state. After a given time delay $t_c$, once the initial frequency comb is restored, the echo process continues to evolve and the stored photons can be readily released to the transmission line. The total storage time is thus given as $1/\Delta+t_c$.

In the experiment, we use the periodicity $\Delta= 3.5\,$MHz and varied close time from $0\,\mu$s to $1.2\,\mu$s. As shown in Fig.~\ref{on-demand}(a), the close stage postpones the revival of the stored photon field, and the echo signal can be effectively extracted once the frequency comb is recovered at the `readout' stage. It is worth noting that the high order echos can be suppress by applying another `close' stage after the first echo. We can therefore on-demand retrieve the stored signal by appropriately setting the close stage. It can be seen that the echo signal gets weaker with an increasing close time. In Fig.~\ref{on-demand}(b), the normalized intensity of the first echo is plotted as a function of storage time, which shows an exponential decay with a time constant of about $0.51\,\mu\,$s. The storage time is mainly limited by the internal loss rate of the resonators, which can be improved by using 3D superconducting resonators \cite{Schoelkopf2016,Kim2019}.

\begin{figure}[htbp]
\centering
\includegraphics[width=1\linewidth]{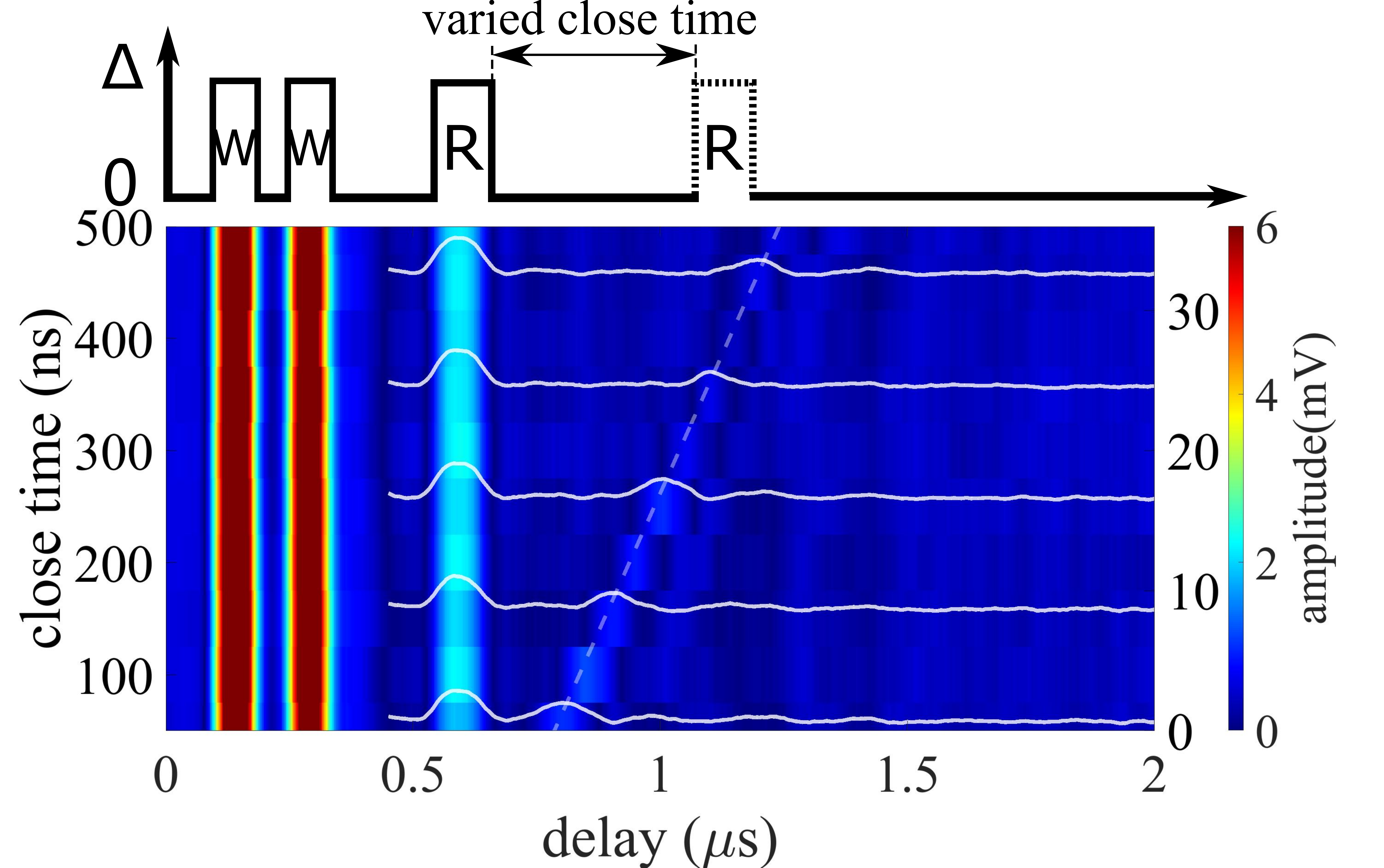}
\caption{\textbf{Multimode storage and on-demand retrieval.} Two input modes, which are defined as two Gaussian shaped input pulses separated by $150\,$ns with FWHM of $50\,$ns, are sent to the multiresonator. The two modes can be separately extracted from the memory, which is manifested by phase correlation between input pulses and echo signals. On demand retrieval of the two modes is manifested by using two close stage of the memory, with the first close time fixed as $455\,$ns for the first mode, and a varied second stage close time ranging from $510\,$ns to $950\,$ns for the second mode. The white solid lines are crosscuts showing the echo intensity as a function of delay. The white dashed line is a guide to the eye indicating the echo signal of the second input mode. The top panel shows sequence of the frequency comb setting, where `W' and `R' represent the write stage and the readout stage, respectively.}
\label{multimode}
\end{figure}


Similar as AFC, the multiresonator device is capable of storing multiple input modes. Considering two temporally distinguishable modes separated by $\delta t$ being sent to the multiresonator, the two photon modes are absorbed sequentially by the frequency comb, and have the same time evolution only with different starting times. Ideally, the echo signals of the two modes would refocused separately with an interval of $\delta t$. Applying the close stage would `freeze' both of the two stored modes, and thus possibly achieving on-demand retrieval of a multimode input signal.
Remarkably, the maximum number of modes that can be stored in the memory depends on the number of the resonators~\cite{Afzelius2009,Moiseev2017}.

To characterize the performance of multimode storage, we send two identical Gaussian-shaped coherent pulses, both with FWHM of $50\,$ns, to the transmission line and measure transmission from the multiresonator. The storage and retrieval flux sequence is illustrated in the top panel of Fig.~\ref{multimode}. The multiresonator is set to a frequency comb with $\Delta = 3.5\,$MHz to capture the input pulses or release the stored photons. Other than that the resonators are always tuned to resonance. We use different storage time for the two modes, with a fixed storage time of $450\,$ns for the first mode, and a storage time varying from $510\,$ns to $950\,$ns for the second mode. As shown in Fig.~\ref{multimode}, the two stored modes can be separately retrieved in accordance with the flux sequence, with a fixed echo time for the first input mode and a controllably delayed echo time for the second mode.


Our device can be used to store a time-bin flying qubit, taking the form $\ket{\psi} = a_e\ket{e}+a_le^{i\phi}\ket{l}$, where $\ket{e}$ and $\ket{l}$ are weak coherent states at the early and late time bin, respectively~\cite{Gisin2008,Hugues2015,Guo2020}. $\phi$ is the relative phase between the two coherent states and $a_e^2+a_l^2=1$. The time-bin qubit is known to be particularly robust to propagation losses, which is a major issue for microwave circuits. We take $a_e = a_l = \sqrt{2}/2$ and vary $\phi$ from $0$ to $2\pi$ to define the input quantum states. We measure the amplitude and relative phase between echo signals of $\ket{e}$ and $\ket{l}$ to characterize the output state.

The fidelity of $\ket{e}$ and $\ket{l}$ can be defined as $F_{e(l)}=(S+N)/(S+2N)$, where $N$ represents the noise and $S$ is the signal strength excluding the noise~\cite{Hugues2015}. The noise mainly comes from non-ideal signal leakage and can be determined by only sending in $\ket{e}$ and measuring signal strength at the expected echo position of $\ket{l}$, and vice versa. By choosing an appropriate storage time, the fidelity of $\ket{e}$ and $\ket{l}$ can be above $99\%$.
For the current setup we are not managing to perform a coherence measurement on the output state~\cite{Gisin2008,Hugues2015}. Instead, we measure the relative amplitude and phase between echo signals of $\ket{e}$ and $\ket{l}$ with varied storage time, which yields a deviation from the ideal values less than $10\%$, showing a good preservation of the input state. Detailed results can be found in Supplementary Information.


To sum up, we have realized on-demand storage and retrieval of microwave photons at the single photon level with a set of frequency-tunable CPW resonators with an overall storage efficiency up to $12\%$.
The storage bandwidth is determined by the spectral distribution of the resonator frequencies, which can reach $55\,$MHz with a trade-off of the storage efficiency. By dynamically tuning the resonator frequencies, microwave photons can be stored in the resonators for a given time and later retrieved on demand. We have also shown that our device is capable of storing more than one input modes, which directly enables on-demand storage and readout of flying quantum states defined by a time-bin qubit. The chip-based device can be directly incorporated into the quantum computing architectures based on superconducting circuits.

This work was supported by  the National Natural Science Foundation of China under Grant No.11874235, the National key Research and Development Program of China (2016YFA0301902), the Frontier Science Center for Quantum Information of the Ministry of Education of China, and the Tsinghua University Initiative Scientific Research Program. Y.K.W. acknowledges support from Shuimu Tsinghua Scholar Program and the International Postdoctoral Exchange Fellowship Program.


\begin{thebibliography}{0}%
\makeatletter
\providecommand \@ifxundefined [1]{%
 \@ifx{#1\undefined}
}%
\providecommand \@ifnum [1]{%
 \ifnum #1\expandafter \@firstoftwo
 \else \expandafter \@secondoftwo
 \fi
}%
\providecommand \@ifx [1]{%
 \ifx #1\expandafter \@firstoftwo
 \else \expandafter \@secondoftwo
 \fi
}%
\providecommand \natexlab [1]{#1}%
\providecommand \enquote  [1]{``#1''}%
\providecommand \bibnamefont  [1]{#1}%
\providecommand \bibfnamefont [1]{#1}%
\providecommand \citenamefont [1]{#1}%
\providecommand \href@noop [0]{\@secondoftwo}%
\providecommand \href [0]{\begingroup \@sanitize@url \@href}%
\providecommand \@href[1]{\@@startlink{#1}\@@href}%
\providecommand \@@href[1]{\endgroup#1\@@endlink}%
\providecommand \@sanitize@url [0]{\catcode `\\12\catcode `\$12\catcode
  `\&12\catcode `\#12\catcode `\^12\catcode `\_12\catcode `\%12\relax}%
\providecommand \@@startlink[1]{}%
\providecommand \@@endlink[0]{}%
\providecommand \url  [0]{\begingroup\@sanitize@url \@url }%
\providecommand \@url [1]{\endgroup\@href {#1}{\urlprefix }}%
\providecommand \urlprefix  [0]{URL }%
\providecommand \Eprint [0]{\href }%
\providecommand \doibase [0]{https://doi.org/}%
\providecommand \selectlanguage [0]{\@gobble}%
\providecommand \bibinfo  [0]{\@secondoftwo}%
\providecommand \bibfield  [0]{\@secondoftwo}%
\providecommand \translation [1]{[#1]}%
\providecommand \BibitemOpen [0]{}%
\providecommand \bibitemStop [0]{}%
\providecommand \bibitemNoStop [0]{.\EOS\space}%
\providecommand \EOS [0]{\spacefactor3000\relax}%
\providecommand \BibitemShut  [1]{\csname bibitem#1\endcsname}%
\let\auto@bib@innerbib\@empty
\end{thebibliography}%


\begin{thebibliography}{41}%
\makeatletter
\providecommand \@ifxundefined [1]{%
 \@ifx{#1\undefined}
}%
\providecommand \@ifnum [1]{%
 \ifnum #1\expandafter \@firstoftwo
 \else \expandafter \@secondoftwo
 \fi
}%
\providecommand \@ifx [1]{%
 \ifx #1\expandafter \@firstoftwo
 \else \expandafter \@secondoftwo
 \fi
}%
\providecommand \natexlab [1]{#1}%
\providecommand \enquote  [1]{``#1''}%
\providecommand \bibnamefont  [1]{#1}%
\providecommand \bibfnamefont [1]{#1}%
\providecommand \citenamefont [1]{#1}%
\providecommand \href@noop [0]{\@secondoftwo}%
\providecommand \href [0]{\begingroup \@sanitize@url \@href}%
\providecommand \@href[1]{\@@startlink{#1}\@@href}%
\providecommand \@@href[1]{\endgroup#1\@@endlink}%
\providecommand \@sanitize@url [0]{\catcode `\\12\catcode `\$12\catcode
  `\&12\catcode `\#12\catcode `\^12\catcode `\_12\catcode `\%12\relax}%
\providecommand \@@startlink[1]{}%
\providecommand \@@endlink[0]{}%
\providecommand \url  [0]{\begingroup\@sanitize@url \@url }%
\providecommand \@url [1]{\endgroup\@href {#1}{\urlprefix }}%
\providecommand \urlprefix  [0]{URL }%
\providecommand \Eprint [0]{\href }%
\providecommand \doibase [0]{https://doi.org/}%
\providecommand \selectlanguage [0]{\@gobble}%
\providecommand \bibinfo  [0]{\@secondoftwo}%
\providecommand \bibfield  [0]{\@secondoftwo}%
\providecommand \translation [1]{[#1]}%
\providecommand \BibitemOpen [0]{}%
\providecommand \bibitemStop [0]{}%
\providecommand \bibitemNoStop [0]{.\EOS\space}%
\providecommand \EOS [0]{\spacefactor3000\relax}%
\providecommand \BibitemShut  [1]{\csname bibitem#1\endcsname}%
\let\auto@bib@innerbib\@empty
\bibitem [{\citenamefont {Kimble}(2008)}]{Kimble2008}%
  \BibitemOpen
  \bibfield  {author} {\bibinfo {author} {\bibfnamefont {H.~J.}\ \bibnamefont
  {Kimble}},\ }\bibfield  {title} {\bibinfo {title} {The quantum internet},\
  }\href@noop {} {\bibfield  {journal} {\bibinfo  {journal} {Nature}\ }\textbf
  {\bibinfo {volume} {453}},\ \bibinfo {pages} {1023} (\bibinfo {year}
  {2008})}\BibitemShut {NoStop}%
\bibitem [{\citenamefont {Lvovsky}\ \emph {et~al.}(2009)\citenamefont
  {Lvovsky}, \citenamefont {Sanders},\ and\ \citenamefont
  {Tittel}}]{Wolfgang2009}%
  \BibitemOpen
  \bibfield  {author} {\bibinfo {author} {\bibfnamefont {A.~I.}\ \bibnamefont
  {Lvovsky}}, \bibinfo {author} {\bibfnamefont {B.~C.}\ \bibnamefont
  {Sanders}},\ and\ \bibinfo {author} {\bibfnamefont {W.}~\bibnamefont
  {Tittel}},\ }\bibfield  {title} {\bibinfo {title} {Optical quantum memory},\
  }\href {https://doi.org/10.1038/nphoton.2009.231} {\bibfield  {journal}
  {\bibinfo  {journal} {Nature Photonics}\ }\textbf {\bibinfo {volume} {3}},\
  \bibinfo {pages} {706} (\bibinfo {year} {2009})}\BibitemShut {NoStop}%
\bibitem [{\citenamefont {Pu}\ \emph {et~al.}(2017)\citenamefont {Pu},
  \citenamefont {Jiang}, \citenamefont {Chang}, \citenamefont {Yang},
  \citenamefont {Li},\ and\ \citenamefont {Duan}}]{Duan2017}%
  \BibitemOpen
  \bibfield  {author} {\bibinfo {author} {\bibfnamefont {Y.~F.}\ \bibnamefont
  {Pu}}, \bibinfo {author} {\bibfnamefont {N.}~\bibnamefont {Jiang}}, \bibinfo
  {author} {\bibfnamefont {W.}~\bibnamefont {Chang}}, \bibinfo {author}
  {\bibfnamefont {H.~X.}\ \bibnamefont {Yang}}, \bibinfo {author}
  {\bibfnamefont {C.}~\bibnamefont {Li}},\ and\ \bibinfo {author}
  {\bibfnamefont {L.~M.}\ \bibnamefont {Duan}},\ }\bibfield  {title} {\bibinfo
  {title} {Experimental realization of a multiplexed quantum memory with 225
  individually accessible memory cells},\ }\href@noop {} {\bibfield  {journal}
  {\bibinfo  {journal} {Nature Communications}\ }\textbf {\bibinfo {volume}
  {8}},\ \bibinfo {pages} {15359} (\bibinfo {year} {2017})}\BibitemShut
  {NoStop}%
\bibitem [{\citenamefont {Clausen}\ \emph {et~al.}(2011)\citenamefont
  {Clausen}, \citenamefont {Usmani}, \citenamefont {Bussières}, \citenamefont
  {Sangouard}, \citenamefont {Afzelius}, \citenamefont {de~Riedmatten},\ and\
  \citenamefont {Gisin}}]{Gisin2011}%
  \BibitemOpen
  \bibfield  {author} {\bibinfo {author} {\bibfnamefont {C.}~\bibnamefont
  {Clausen}}, \bibinfo {author} {\bibfnamefont {I.}~\bibnamefont {Usmani}},
  \bibinfo {author} {\bibfnamefont {F.}~\bibnamefont {Bussières}}, \bibinfo
  {author} {\bibfnamefont {N.}~\bibnamefont {Sangouard}}, \bibinfo {author}
  {\bibfnamefont {M.}~\bibnamefont {Afzelius}}, \bibinfo {author}
  {\bibfnamefont {H.}~\bibnamefont {de~Riedmatten}},\ and\ \bibinfo {author}
  {\bibfnamefont {N.}~\bibnamefont {Gisin}},\ }\bibfield  {title} {\bibinfo
  {title} {Quantum storage of photonic entanglement in a crystal},\ }\href@noop
  {} {\bibfield  {journal} {\bibinfo  {journal} {Nature}\ }\textbf {\bibinfo
  {volume} {469}},\ \bibinfo {pages} {508} (\bibinfo {year}
  {2011})}\BibitemShut {NoStop}%
\bibitem [{\citenamefont {Saglamyurek}\ \emph {et~al.}(2011)\citenamefont
  {Saglamyurek}, \citenamefont {Sinclair}, \citenamefont {Jin}, \citenamefont
  {Slater}, \citenamefont {Oblak}, \citenamefont {Bussières}, \citenamefont
  {George}, \citenamefont {Ricken}, \citenamefont {Sohler},\ and\ \citenamefont
  {Tittel}}]{Tittel2011}%
  \BibitemOpen
  \bibfield  {author} {\bibinfo {author} {\bibfnamefont {E.}~\bibnamefont
  {Saglamyurek}}, \bibinfo {author} {\bibfnamefont {N.}~\bibnamefont
  {Sinclair}}, \bibinfo {author} {\bibfnamefont {J.}~\bibnamefont {Jin}},
  \bibinfo {author} {\bibfnamefont {J.~A.}\ \bibnamefont {Slater}}, \bibinfo
  {author} {\bibfnamefont {D.}~\bibnamefont {Oblak}}, \bibinfo {author}
  {\bibfnamefont {F.}~\bibnamefont {Bussières}}, \bibinfo {author}
  {\bibfnamefont {M.}~\bibnamefont {George}}, \bibinfo {author} {\bibfnamefont
  {R.}~\bibnamefont {Ricken}}, \bibinfo {author} {\bibfnamefont
  {W.}~\bibnamefont {Sohler}},\ and\ \bibinfo {author} {\bibfnamefont
  {W.}~\bibnamefont {Tittel}},\ }\bibfield  {title} {\bibinfo {title}
  {Broadband waveguide quantum memory for entangled photons},\ }\href@noop {}
  {\bibfield  {journal} {\bibinfo  {journal} {Nature}\ }\textbf {\bibinfo
  {volume} {469}},\ \bibinfo {pages} {512} (\bibinfo {year}
  {2011})}\BibitemShut {NoStop}%
\bibitem [{\citenamefont {Lobino}\ \emph {et~al.}(2009)\citenamefont {Lobino},
  \citenamefont {Kupchak}, \citenamefont {Figueroa},\ and\ \citenamefont
  {Lvovsky}}]{Lvovsky2009}%
  \BibitemOpen
  \bibfield  {author} {\bibinfo {author} {\bibfnamefont {M.}~\bibnamefont
  {Lobino}}, \bibinfo {author} {\bibfnamefont {C.}~\bibnamefont {Kupchak}},
  \bibinfo {author} {\bibfnamefont {E.}~\bibnamefont {Figueroa}},\ and\
  \bibinfo {author} {\bibfnamefont {A.~I.}\ \bibnamefont {Lvovsky}},\
  }\bibfield  {title} {\bibinfo {title} {Memory for light as a quantum
  process},\ }\href@noop {} {\bibfield  {journal} {\bibinfo  {journal} {Phys.
  Rev. Lett.}\ }\textbf {\bibinfo {volume} {102}},\ \bibinfo {pages} {203601}
  (\bibinfo {year} {2009})}\BibitemShut {NoStop}%
\bibitem [{\citenamefont {Hedges}\ \emph {et~al.}(2010)\citenamefont {Hedges},
  \citenamefont {Longdell}, \citenamefont {Li},\ and\ \citenamefont
  {Sellars}}]{Sellars2010}%
  \BibitemOpen
  \bibfield  {author} {\bibinfo {author} {\bibfnamefont {M.~P.}\ \bibnamefont
  {Hedges}}, \bibinfo {author} {\bibfnamefont {J.~J.}\ \bibnamefont
  {Longdell}}, \bibinfo {author} {\bibfnamefont {Y.}~\bibnamefont {Li}},\ and\
  \bibinfo {author} {\bibfnamefont {M.~J.}\ \bibnamefont {Sellars}},\
  }\bibfield  {title} {\bibinfo {title} {Efficient quantum memory for light},\
  }\href@noop {} {\bibfield  {journal} {\bibinfo  {journal} {Nature}\ }\textbf
  {\bibinfo {volume} {465}},\ \bibinfo {pages} {1052} (\bibinfo {year}
  {2010})}\BibitemShut {NoStop}%
\bibitem [{\citenamefont {Riel\"ander}\ \emph {et~al.}(2014)\citenamefont
  {Riel\"ander}, \citenamefont {Kutluer}, \citenamefont {Ledingham},
  \citenamefont {G\"undo\ifmmode~\breve{g}\else \u{g}\fi{}an}, \citenamefont
  {Fekete}, \citenamefont {Mazzera},\ and\ \citenamefont
  {de~Riedmatten}}]{Hugues2014}%
  \BibitemOpen
  \bibfield  {author} {\bibinfo {author} {\bibfnamefont {D.}~\bibnamefont
  {Riel\"ander}}, \bibinfo {author} {\bibfnamefont {K.}~\bibnamefont
  {Kutluer}}, \bibinfo {author} {\bibfnamefont {P.~M.}\ \bibnamefont
  {Ledingham}}, \bibinfo {author} {\bibfnamefont {M.}~\bibnamefont
  {G\"undo\ifmmode~\breve{g}\else \u{g}\fi{}an}}, \bibinfo {author}
  {\bibfnamefont {J.}~\bibnamefont {Fekete}}, \bibinfo {author} {\bibfnamefont
  {M.}~\bibnamefont {Mazzera}},\ and\ \bibinfo {author} {\bibfnamefont
  {H.}~\bibnamefont {de~Riedmatten}},\ }\bibfield  {title} {\bibinfo {title}
  {Quantum storage of heralded single photons in a praseodymium-doped
  crystal},\ }\href@noop {} {\bibfield  {journal} {\bibinfo  {journal} {Phys.
  Rev. Lett.}\ }\textbf {\bibinfo {volume} {112}},\ \bibinfo {pages} {040504}
  (\bibinfo {year} {2014})}\BibitemShut {NoStop}%
\bibitem [{\citenamefont {Tang}\ \emph {et~al.}(2015)\citenamefont {Tang},
  \citenamefont {Zhou}, \citenamefont {Wang}, \citenamefont {Li}, \citenamefont
  {Liu}, \citenamefont {Hua}, \citenamefont {Zou}, \citenamefont {Wang},
  \citenamefont {He}, \citenamefont {Chen}, \citenamefont {Sun}, \citenamefont
  {Yu}, \citenamefont {Li}, \citenamefont {Zha}, \citenamefont {Ni},
  \citenamefont {Niu}, \citenamefont {Li},\ and\ \citenamefont
  {Guo}}]{Guo2015}%
  \BibitemOpen
  \bibfield  {author} {\bibinfo {author} {\bibfnamefont {J.-S.}\ \bibnamefont
  {Tang}}, \bibinfo {author} {\bibfnamefont {Z.-Q.}\ \bibnamefont {Zhou}},
  \bibinfo {author} {\bibfnamefont {Y.-T.}\ \bibnamefont {Wang}}, \bibinfo
  {author} {\bibfnamefont {Y.-L.}\ \bibnamefont {Li}}, \bibinfo {author}
  {\bibfnamefont {X.}~\bibnamefont {Liu}}, \bibinfo {author} {\bibfnamefont
  {Y.-L.}\ \bibnamefont {Hua}}, \bibinfo {author} {\bibfnamefont
  {Y.}~\bibnamefont {Zou}}, \bibinfo {author} {\bibfnamefont {S.}~\bibnamefont
  {Wang}}, \bibinfo {author} {\bibfnamefont {D.-Y.}\ \bibnamefont {He}},
  \bibinfo {author} {\bibfnamefont {G.}~\bibnamefont {Chen}}, \bibinfo {author}
  {\bibfnamefont {Y.-N.}\ \bibnamefont {Sun}}, \bibinfo {author} {\bibfnamefont
  {Y.}~\bibnamefont {Yu}}, \bibinfo {author} {\bibfnamefont {M.-F.}\
  \bibnamefont {Li}}, \bibinfo {author} {\bibfnamefont {G.-W.}\ \bibnamefont
  {Zha}}, \bibinfo {author} {\bibfnamefont {H.-Q.}\ \bibnamefont {Ni}},
  \bibinfo {author} {\bibfnamefont {Z.-C.}\ \bibnamefont {Niu}}, \bibinfo
  {author} {\bibfnamefont {C.-F.}\ \bibnamefont {Li}},\ and\ \bibinfo {author}
  {\bibfnamefont {G.-C.}\ \bibnamefont {Guo}},\ }\bibfield  {title} {\bibinfo
  {title} {Storage of multiple single-photon pulses emitted from a quantum dot
  in a solid-state quantum memory},\ }\href@noop {} {\bibfield  {journal}
  {\bibinfo  {journal} {Nature Communications}\ }\textbf {\bibinfo {volume}
  {6}},\ \bibinfo {pages} {8652} (\bibinfo {year} {2015})}\BibitemShut
  {NoStop}%
\bibitem [{\citenamefont {Bouillard}\ \emph {et~al.}(2019)\citenamefont
  {Bouillard}, \citenamefont {Boucher}, \citenamefont {Ferrer~Ortas},
  \citenamefont {Pointard},\ and\ \citenamefont {Tualle-Brouri}}]{Tualle2019}%
  \BibitemOpen
  \bibfield  {author} {\bibinfo {author} {\bibfnamefont {M.}~\bibnamefont
  {Bouillard}}, \bibinfo {author} {\bibfnamefont {G.}~\bibnamefont {Boucher}},
  \bibinfo {author} {\bibfnamefont {J.}~\bibnamefont {Ferrer~Ortas}}, \bibinfo
  {author} {\bibfnamefont {B.}~\bibnamefont {Pointard}},\ and\ \bibinfo
  {author} {\bibfnamefont {R.}~\bibnamefont {Tualle-Brouri}},\ }\bibfield
  {title} {\bibinfo {title} {Quantum storage of single-photon and two-photon
  fock states with an all-optical quantum memory},\ }\href@noop {} {\bibfield
  {journal} {\bibinfo  {journal} {Phys. Rev. Lett.}\ }\textbf {\bibinfo
  {volume} {122}},\ \bibinfo {pages} {210501} (\bibinfo {year}
  {2019})}\BibitemShut {NoStop}%
\bibitem [{\citenamefont {Zhou}\ \emph {et~al.}(2012)\citenamefont {Zhou},
  \citenamefont {Lin}, \citenamefont {Yang}, \citenamefont {Li},\ and\
  \citenamefont {Guo}}]{Guo2012}%
  \BibitemOpen
  \bibfield  {author} {\bibinfo {author} {\bibfnamefont {Z.-Q.}\ \bibnamefont
  {Zhou}}, \bibinfo {author} {\bibfnamefont {W.-B.}\ \bibnamefont {Lin}},
  \bibinfo {author} {\bibfnamefont {M.}~\bibnamefont {Yang}}, \bibinfo {author}
  {\bibfnamefont {C.-F.}\ \bibnamefont {Li}},\ and\ \bibinfo {author}
  {\bibfnamefont {G.-C.}\ \bibnamefont {Guo}},\ }\bibfield  {title} {\bibinfo
  {title} {Realization of reliable solid-state quantum memory for photonic
  polarization qubit},\ }\href {https://doi.org/10.1103/PhysRevLett.108.190505}
  {\bibfield  {journal} {\bibinfo  {journal} {Phys. Rev. Lett.}\ }\textbf
  {\bibinfo {volume} {108}},\ \bibinfo {pages} {190505} (\bibinfo {year}
  {2012})}\BibitemShut {NoStop}%
\bibitem [{\citenamefont {Wang}\ \emph {et~al.}(2019)\citenamefont {Wang},
  \citenamefont {Li}, \citenamefont {Zhang}, \citenamefont {Su}, \citenamefont
  {Zhou}, \citenamefont {Liao}, \citenamefont {Du}, \citenamefont {Yan},\ and\
  \citenamefont {Zhu}}]{Zhu2019}%
  \BibitemOpen
  \bibfield  {author} {\bibinfo {author} {\bibfnamefont {Y.}~\bibnamefont
  {Wang}}, \bibinfo {author} {\bibfnamefont {J.}~\bibnamefont {Li}}, \bibinfo
  {author} {\bibfnamefont {S.}~\bibnamefont {Zhang}}, \bibinfo {author}
  {\bibfnamefont {K.}~\bibnamefont {Su}}, \bibinfo {author} {\bibfnamefont
  {Y.}~\bibnamefont {Zhou}}, \bibinfo {author} {\bibfnamefont {K.}~\bibnamefont
  {Liao}}, \bibinfo {author} {\bibfnamefont {S.}~\bibnamefont {Du}}, \bibinfo
  {author} {\bibfnamefont {H.}~\bibnamefont {Yan}},\ and\ \bibinfo {author}
  {\bibfnamefont {S.-L.}\ \bibnamefont {Zhu}},\ }\bibfield  {title} {\bibinfo
  {title} {Efficient quantum memory for single-photon polarization qubits},\
  }\href@noop {} {\bibfield  {journal} {\bibinfo  {journal} {Nature Photonics}\
  }\textbf {\bibinfo {volume} {13}},\ \bibinfo {pages} {346} (\bibinfo {year}
  {2019})}\BibitemShut {NoStop}%
\bibitem [{\citenamefont {Liu}\ \emph {et~al.}(2020)\citenamefont {Liu},
  \citenamefont {Zhu}, \citenamefont {Su}, \citenamefont {Ma}, \citenamefont
  {Zhou}, \citenamefont {Li},\ and\ \citenamefont {Guo}}]{Guo2020}%
  \BibitemOpen
  \bibfield  {author} {\bibinfo {author} {\bibfnamefont {C.}~\bibnamefont
  {Liu}}, \bibinfo {author} {\bibfnamefont {T.-X.}\ \bibnamefont {Zhu}},
  \bibinfo {author} {\bibfnamefont {M.-X.}\ \bibnamefont {Su}}, \bibinfo
  {author} {\bibfnamefont {Y.-Z.}\ \bibnamefont {Ma}}, \bibinfo {author}
  {\bibfnamefont {Z.-Q.}\ \bibnamefont {Zhou}}, \bibinfo {author}
  {\bibfnamefont {C.-F.}\ \bibnamefont {Li}},\ and\ \bibinfo {author}
  {\bibfnamefont {G.-C.}\ \bibnamefont {Guo}},\ }\bibfield  {title} {\bibinfo
  {title} {On-demand quantum storage of photonic qubits in an on-chip
  waveguide},\ }\href@noop {} {\bibfield  {journal} {\bibinfo  {journal} {Phys.
  Rev. Lett.}\ }\textbf {\bibinfo {volume} {125}},\ \bibinfo {pages} {260504}
  (\bibinfo {year} {2020})}\BibitemShut {NoStop}%
\bibitem [{\citenamefont {Duan}\ \emph {et~al.}(2001)\citenamefont {Duan},
  \citenamefont {Lukin}, \citenamefont {Cirac},\ and\ \citenamefont
  {Zoller}}]{Duan2001}%
  \BibitemOpen
  \bibfield  {author} {\bibinfo {author} {\bibfnamefont {L.~M.}\ \bibnamefont
  {Duan}}, \bibinfo {author} {\bibfnamefont {M.~D.}\ \bibnamefont {Lukin}},
  \bibinfo {author} {\bibfnamefont {J.~I.}\ \bibnamefont {Cirac}},\ and\
  \bibinfo {author} {\bibfnamefont {P.}~\bibnamefont {Zoller}},\ }\bibfield
  {title} {\bibinfo {title} {Long-distance quantum communication with atomic
  ensembles and linear optics},\ }\href@noop {} {\bibfield  {journal} {\bibinfo
   {journal} {Nature}\ }\textbf {\bibinfo {volume} {414}},\ \bibinfo {pages}
  {413} (\bibinfo {year} {2001})}\BibitemShut {NoStop}%
\bibitem [{\citenamefont {Wehner}\ \emph {et~al.}(2018)\citenamefont {Wehner},
  \citenamefont {Elkouss},\ and\ \citenamefont {Hanson}}]{Hanson2018}%
  \BibitemOpen
  \bibfield  {author} {\bibinfo {author} {\bibfnamefont {S.}~\bibnamefont
  {Wehner}}, \bibinfo {author} {\bibfnamefont {D.}~\bibnamefont {Elkouss}},\
  and\ \bibinfo {author} {\bibfnamefont {R.}~\bibnamefont {Hanson}},\
  }\bibfield  {title} {\bibinfo {title} {Quantum internet: A vision for the
  road ahead},\ }\href@noop {} {\bibfield  {journal} {\bibinfo  {journal}
  {Science}\ }\textbf {\bibinfo {volume} {362}} (\bibinfo {year}
  {2018})}\BibitemShut {NoStop}%
\bibitem [{\citenamefont {Devoret}\ and\ \citenamefont
  {Schoelkopf}(2013)}]{Devoret2013}%
  \BibitemOpen
  \bibfield  {author} {\bibinfo {author} {\bibfnamefont {M.~H.}\ \bibnamefont
  {Devoret}}\ and\ \bibinfo {author} {\bibfnamefont {R.~J.}\ \bibnamefont
  {Schoelkopf}},\ }\bibfield  {title} {\bibinfo {title} {Superconducting
  circuits for quantum information: An outlook},\ }\href
  {https://doi.org/10.1126/science.1231930} {\bibfield  {journal} {\bibinfo
  {journal} {Science}\ }\textbf {\bibinfo {volume} {339}},\ \bibinfo {pages}
  {1169} (\bibinfo {year} {2013})}\BibitemShut {NoStop}%
\bibitem [{\citenamefont {Kurizki}\ \emph {et~al.}(2015)\citenamefont
  {Kurizki}, \citenamefont {Bertet}, \citenamefont {Kubo}, \citenamefont
  {M{\o}lmer}, \citenamefont {Petrosyan}, \citenamefont {Rabl},\ and\
  \citenamefont {Schmiedmayer}}]{Kurizki2015}%
  \BibitemOpen
  \bibfield  {author} {\bibinfo {author} {\bibfnamefont {G.}~\bibnamefont
  {Kurizki}}, \bibinfo {author} {\bibfnamefont {P.}~\bibnamefont {Bertet}},
  \bibinfo {author} {\bibfnamefont {Y.}~\bibnamefont {Kubo}}, \bibinfo {author}
  {\bibfnamefont {K.}~\bibnamefont {M{\o}lmer}}, \bibinfo {author}
  {\bibfnamefont {D.}~\bibnamefont {Petrosyan}}, \bibinfo {author}
  {\bibfnamefont {P.}~\bibnamefont {Rabl}},\ and\ \bibinfo {author}
  {\bibfnamefont {J.}~\bibnamefont {Schmiedmayer}},\ }\bibfield  {title}
  {\bibinfo {title} {Quantum technologies with hybrid systems},\ }\href
  {https://doi.org/10.1073/pnas.1419326112} {\bibfield  {journal} {\bibinfo
  {journal} {Proceedings of the National Academy of Sciences}\ }\textbf
  {\bibinfo {volume} {112}},\ \bibinfo {pages} {3866} (\bibinfo {year}
  {2015})}\BibitemShut {NoStop}%
\bibitem [{\citenamefont {Kubo}\ \emph {et~al.}(2011)\citenamefont {Kubo},
  \citenamefont {Grezes}, \citenamefont {Dewes}, \citenamefont {Umeda},
  \citenamefont {Isoya}, \citenamefont {Sumiya}, \citenamefont {Morishita},
  \citenamefont {Abe}, \citenamefont {Onoda}, \citenamefont {Ohshima},
  \citenamefont {Jacques}, \citenamefont {Dr\'eau}, \citenamefont {Roch},
  \citenamefont {Diniz}, \citenamefont {Auffeves}, \citenamefont {Vion},
  \citenamefont {Esteve},\ and\ \citenamefont {Bertet}}]{Kubo2011}%
  \BibitemOpen
  \bibfield  {author} {\bibinfo {author} {\bibfnamefont {Y.}~\bibnamefont
  {Kubo}}, \bibinfo {author} {\bibfnamefont {C.}~\bibnamefont {Grezes}},
  \bibinfo {author} {\bibfnamefont {A.}~\bibnamefont {Dewes}}, \bibinfo
  {author} {\bibfnamefont {T.}~\bibnamefont {Umeda}}, \bibinfo {author}
  {\bibfnamefont {J.}~\bibnamefont {Isoya}}, \bibinfo {author} {\bibfnamefont
  {H.}~\bibnamefont {Sumiya}}, \bibinfo {author} {\bibfnamefont
  {N.}~\bibnamefont {Morishita}}, \bibinfo {author} {\bibfnamefont
  {H.}~\bibnamefont {Abe}}, \bibinfo {author} {\bibfnamefont {S.}~\bibnamefont
  {Onoda}}, \bibinfo {author} {\bibfnamefont {T.}~\bibnamefont {Ohshima}},
  \bibinfo {author} {\bibfnamefont {V.}~\bibnamefont {Jacques}}, \bibinfo
  {author} {\bibfnamefont {A.}~\bibnamefont {Dr\'eau}}, \bibinfo {author}
  {\bibfnamefont {J.-F.}\ \bibnamefont {Roch}}, \bibinfo {author}
  {\bibfnamefont {I.}~\bibnamefont {Diniz}}, \bibinfo {author} {\bibfnamefont
  {A.}~\bibnamefont {Auffeves}}, \bibinfo {author} {\bibfnamefont
  {D.}~\bibnamefont {Vion}}, \bibinfo {author} {\bibfnamefont {D.}~\bibnamefont
  {Esteve}},\ and\ \bibinfo {author} {\bibfnamefont {P.}~\bibnamefont
  {Bertet}},\ }\bibfield  {title} {\bibinfo {title} {Hybrid quantum circuit
  with a superconducting qubit coupled to a spin ensemble},\ }\href
  {https://doi.org/10.1103/PhysRevLett.107.220501} {\bibfield  {journal}
  {\bibinfo  {journal} {Phys. Rev. Lett.}\ }\textbf {\bibinfo {volume} {107}},\
  \bibinfo {pages} {220501} (\bibinfo {year} {2011})}\BibitemShut {NoStop}%
\bibitem [{\citenamefont {Naik}\ \emph {et~al.}(2017)\citenamefont {Naik},
  \citenamefont {Leung}, \citenamefont {Chakram}, \citenamefont {Groszkowski},
  \citenamefont {Lu}, \citenamefont {Earnest}, \citenamefont {McKay},
  \citenamefont {Koch},\ and\ \citenamefont {Schuster}}]{Schuster2017}%
  \BibitemOpen
  \bibfield  {author} {\bibinfo {author} {\bibfnamefont {R.~K.}\ \bibnamefont
  {Naik}}, \bibinfo {author} {\bibfnamefont {N.}~\bibnamefont {Leung}},
  \bibinfo {author} {\bibfnamefont {S.}~\bibnamefont {Chakram}}, \bibinfo
  {author} {\bibfnamefont {P.}~\bibnamefont {Groszkowski}}, \bibinfo {author}
  {\bibfnamefont {Y.}~\bibnamefont {Lu}}, \bibinfo {author} {\bibfnamefont
  {N.}~\bibnamefont {Earnest}}, \bibinfo {author} {\bibfnamefont {D.~C.}\
  \bibnamefont {McKay}}, \bibinfo {author} {\bibfnamefont {J.}~\bibnamefont
  {Koch}},\ and\ \bibinfo {author} {\bibfnamefont {D.~I.}\ \bibnamefont
  {Schuster}},\ }\bibfield  {title} {\bibinfo {title} {Random access quantum
  information processors using multimode circuit quantum electrodynamics},\
  }\href {https://doi.org/10.1038/s41467-017-02046-6} {\bibfield  {journal}
  {\bibinfo  {journal} {Nature Communications}\ }\textbf {\bibinfo {volume}
  {8}},\ \bibinfo {pages} {1904} (\bibinfo {year} {2017})}\BibitemShut
  {NoStop}%
\bibitem [{\citenamefont {Imamo\ifmmode~\breve{g}\else
  \u{g}\fi{}lu}(2009)}]{Imamo2009}%
  \BibitemOpen
  \bibfield  {author} {\bibinfo {author} {\bibfnamefont {A.}~\bibnamefont
  {Imamo\ifmmode~\breve{g}\else \u{g}\fi{}lu}},\ }\bibfield  {title} {\bibinfo
  {title} {Cavity qed based on collective magnetic dipole coupling: Spin
  ensembles as hybrid two-level systems},\ }\href@noop {} {\bibfield  {journal}
  {\bibinfo  {journal} {Phys. Rev. Lett.}\ }\textbf {\bibinfo {volume} {102}},\
  \bibinfo {pages} {083602} (\bibinfo {year} {2009})}\BibitemShut {NoStop}%
\bibitem [{\citenamefont {Zhu}\ \emph {et~al.}(2011)\citenamefont {Zhu},
  \citenamefont {Saito}, \citenamefont {Kemp}, \citenamefont {Kakuyanagi},
  \citenamefont {Karimoto}, \citenamefont {Nakano}, \citenamefont {Munro},
  \citenamefont {Tokura}, \citenamefont {Everitt}, \citenamefont {Nemoto},
  \citenamefont {Kasu}, \citenamefont {Mizuochi},\ and\ \citenamefont
  {Semba}}]{Zhu2011}%
  \BibitemOpen
  \bibfield  {author} {\bibinfo {author} {\bibfnamefont {X.}~\bibnamefont
  {Zhu}}, \bibinfo {author} {\bibfnamefont {S.}~\bibnamefont {Saito}}, \bibinfo
  {author} {\bibfnamefont {A.}~\bibnamefont {Kemp}}, \bibinfo {author}
  {\bibfnamefont {K.}~\bibnamefont {Kakuyanagi}}, \bibinfo {author}
  {\bibfnamefont {S.-i.}\ \bibnamefont {Karimoto}}, \bibinfo {author}
  {\bibfnamefont {H.}~\bibnamefont {Nakano}}, \bibinfo {author} {\bibfnamefont
  {W.~J.}\ \bibnamefont {Munro}}, \bibinfo {author} {\bibfnamefont
  {Y.}~\bibnamefont {Tokura}}, \bibinfo {author} {\bibfnamefont {M.~S.}\
  \bibnamefont {Everitt}}, \bibinfo {author} {\bibfnamefont {K.}~\bibnamefont
  {Nemoto}}, \bibinfo {author} {\bibfnamefont {M.}~\bibnamefont {Kasu}},
  \bibinfo {author} {\bibfnamefont {N.}~\bibnamefont {Mizuochi}},\ and\
  \bibinfo {author} {\bibfnamefont {K.}~\bibnamefont {Semba}},\ }\bibfield
  {title} {\bibinfo {title} {Coherent coupling of a superconducting flux qubit
  to an electron spin ensemble in diamond},\ }\href
  {https://doi.org/10.1038/nature10462} {\bibfield  {journal} {\bibinfo
  {journal} {Nature}\ }\textbf {\bibinfo {volume} {478}},\ \bibinfo {pages}
  {221} (\bibinfo {year} {2011})}\BibitemShut {NoStop}%
\bibitem [{\citenamefont {Ams\"uss}\ \emph {et~al.}(2011)\citenamefont
  {Ams\"uss}, \citenamefont {Koller}, \citenamefont {N\"obauer}, \citenamefont
  {Putz}, \citenamefont {Rotter}, \citenamefont {Sandner}, \citenamefont
  {Schneider}, \citenamefont {Schramb\"ock}, \citenamefont {Steinhauser},
  \citenamefont {Ritsch}, \citenamefont {Schmiedmayer},\ and\ \citenamefont
  {Majer}}]{Majer2011}%
  \BibitemOpen
  \bibfield  {author} {\bibinfo {author} {\bibfnamefont {R.}~\bibnamefont
  {Ams\"uss}}, \bibinfo {author} {\bibfnamefont {C.}~\bibnamefont {Koller}},
  \bibinfo {author} {\bibfnamefont {T.}~\bibnamefont {N\"obauer}}, \bibinfo
  {author} {\bibfnamefont {S.}~\bibnamefont {Putz}}, \bibinfo {author}
  {\bibfnamefont {S.}~\bibnamefont {Rotter}}, \bibinfo {author} {\bibfnamefont
  {K.}~\bibnamefont {Sandner}}, \bibinfo {author} {\bibfnamefont
  {S.}~\bibnamefont {Schneider}}, \bibinfo {author} {\bibfnamefont
  {M.}~\bibnamefont {Schramb\"ock}}, \bibinfo {author} {\bibfnamefont
  {G.}~\bibnamefont {Steinhauser}}, \bibinfo {author} {\bibfnamefont
  {H.}~\bibnamefont {Ritsch}}, \bibinfo {author} {\bibfnamefont
  {J.}~\bibnamefont {Schmiedmayer}},\ and\ \bibinfo {author} {\bibfnamefont
  {J.}~\bibnamefont {Majer}},\ }\bibfield  {title} {\bibinfo {title} {Cavity
  qed with magnetically coupled collective spin states},\ }\href@noop {}
  {\bibfield  {journal} {\bibinfo  {journal} {Phys. Rev. Lett.}\ }\textbf
  {\bibinfo {volume} {107}},\ \bibinfo {pages} {060502} (\bibinfo {year}
  {2011})}\BibitemShut {NoStop}%
\bibitem [{\citenamefont {Julsgaard}\ \emph {et~al.}(2013)\citenamefont
  {Julsgaard}, \citenamefont {Grezes}, \citenamefont {Bertet},\ and\
  \citenamefont {M\o{}lmer}}]{Julsgaard2013}%
  \BibitemOpen
  \bibfield  {author} {\bibinfo {author} {\bibfnamefont {B.}~\bibnamefont
  {Julsgaard}}, \bibinfo {author} {\bibfnamefont {C.}~\bibnamefont {Grezes}},
  \bibinfo {author} {\bibfnamefont {P.}~\bibnamefont {Bertet}},\ and\ \bibinfo
  {author} {\bibfnamefont {K.}~\bibnamefont {M\o{}lmer}},\ }\bibfield  {title}
  {\bibinfo {title} {Quantum memory for microwave photons in an inhomogeneously
  broadened spin ensemble},\ }\href
  {https://doi.org/10.1103/PhysRevLett.110.250503} {\bibfield  {journal}
  {\bibinfo  {journal} {Phys. Rev. Lett.}\ }\textbf {\bibinfo {volume} {110}},\
  \bibinfo {pages} {250503} (\bibinfo {year} {2013})}\BibitemShut {NoStop}%
\bibitem [{\citenamefont {Zhang}\ \emph {et~al.}(2015)\citenamefont {Zhang},
  \citenamefont {Zou}, \citenamefont {Zhu}, \citenamefont {Marquardt},
  \citenamefont {Jiang},\ and\ \citenamefont {Tang}}]{Jiang2015}%
  \BibitemOpen
  \bibfield  {author} {\bibinfo {author} {\bibfnamefont {X.}~\bibnamefont
  {Zhang}}, \bibinfo {author} {\bibfnamefont {C.-L.}\ \bibnamefont {Zou}},
  \bibinfo {author} {\bibfnamefont {N.}~\bibnamefont {Zhu}}, \bibinfo {author}
  {\bibfnamefont {F.}~\bibnamefont {Marquardt}}, \bibinfo {author}
  {\bibfnamefont {L.}~\bibnamefont {Jiang}},\ and\ \bibinfo {author}
  {\bibfnamefont {H.~X.}\ \bibnamefont {Tang}},\ }\bibfield  {title} {\bibinfo
  {title} {Magnon dark modes and gradient memory},\ }\href
  {https://doi.org/10.1038/ncomms9914} {\bibfield  {journal} {\bibinfo
  {journal} {Nature Communications}\ }\textbf {\bibinfo {volume} {6}},\
  \bibinfo {pages} {8914} (\bibinfo {year} {2015})}\BibitemShut {NoStop}%
\bibitem [{\citenamefont {Ranjan}\ \emph {et~al.}(2020)\citenamefont {Ranjan},
  \citenamefont {O'Sullivan}, \citenamefont {Albertinale}, \citenamefont
  {Albanese}, \citenamefont {Chaneli\`ere}, \citenamefont {Schenkel},
  \citenamefont {Vion}, \citenamefont {Esteve}, \citenamefont {Flurin},
  \citenamefont {Morton},\ and\ \citenamefont {Bertet}}]{Bertet2020}%
  \BibitemOpen
  \bibfield  {author} {\bibinfo {author} {\bibfnamefont {V.}~\bibnamefont
  {Ranjan}}, \bibinfo {author} {\bibfnamefont {J.}~\bibnamefont {O'Sullivan}},
  \bibinfo {author} {\bibfnamefont {E.}~\bibnamefont {Albertinale}}, \bibinfo
  {author} {\bibfnamefont {B.}~\bibnamefont {Albanese}}, \bibinfo {author}
  {\bibfnamefont {T.}~\bibnamefont {Chaneli\`ere}}, \bibinfo {author}
  {\bibfnamefont {T.}~\bibnamefont {Schenkel}}, \bibinfo {author}
  {\bibfnamefont {D.}~\bibnamefont {Vion}}, \bibinfo {author} {\bibfnamefont
  {D.}~\bibnamefont {Esteve}}, \bibinfo {author} {\bibfnamefont
  {E.}~\bibnamefont {Flurin}}, \bibinfo {author} {\bibfnamefont {J.~J.~L.}\
  \bibnamefont {Morton}},\ and\ \bibinfo {author} {\bibfnamefont
  {P.}~\bibnamefont {Bertet}},\ }\bibfield  {title} {\bibinfo {title}
  {Multimode storage of quantum microwave fields in electron spins over 100
  ms},\ }\href@noop {} {\bibfield  {journal} {\bibinfo  {journal} {Phys. Rev.
  Lett.}\ }\textbf {\bibinfo {volume} {125}},\ \bibinfo {pages} {210505}
  (\bibinfo {year} {2020})}\BibitemShut {NoStop}%
\bibitem [{\citenamefont {Grezes}\ \emph {et~al.}(2014)\citenamefont {Grezes},
  \citenamefont {Julsgaard}, \citenamefont {Kubo}, \citenamefont {Stern},
  \citenamefont {Umeda}, \citenamefont {Isoya}, \citenamefont {Sumiya},
  \citenamefont {Abe}, \citenamefont {Onoda}, \citenamefont {Ohshima},
  \citenamefont {Jacques}, \citenamefont {Esteve}, \citenamefont {Vion},
  \citenamefont {Esteve}, \citenamefont {M\o{}lmer},\ and\ \citenamefont
  {Bertet}}]{Grezes2014}%
  \BibitemOpen
  \bibfield  {author} {\bibinfo {author} {\bibfnamefont {C.}~\bibnamefont
  {Grezes}}, \bibinfo {author} {\bibfnamefont {B.}~\bibnamefont {Julsgaard}},
  \bibinfo {author} {\bibfnamefont {Y.}~\bibnamefont {Kubo}}, \bibinfo {author}
  {\bibfnamefont {M.}~\bibnamefont {Stern}}, \bibinfo {author} {\bibfnamefont
  {T.}~\bibnamefont {Umeda}}, \bibinfo {author} {\bibfnamefont
  {J.}~\bibnamefont {Isoya}}, \bibinfo {author} {\bibfnamefont
  {H.}~\bibnamefont {Sumiya}}, \bibinfo {author} {\bibfnamefont
  {H.}~\bibnamefont {Abe}}, \bibinfo {author} {\bibfnamefont {S.}~\bibnamefont
  {Onoda}}, \bibinfo {author} {\bibfnamefont {T.}~\bibnamefont {Ohshima}},
  \bibinfo {author} {\bibfnamefont {V.}~\bibnamefont {Jacques}}, \bibinfo
  {author} {\bibfnamefont {J.}~\bibnamefont {Esteve}}, \bibinfo {author}
  {\bibfnamefont {D.}~\bibnamefont {Vion}}, \bibinfo {author} {\bibfnamefont
  {D.}~\bibnamefont {Esteve}}, \bibinfo {author} {\bibfnamefont
  {K.}~\bibnamefont {M\o{}lmer}},\ and\ \bibinfo {author} {\bibfnamefont
  {P.}~\bibnamefont {Bertet}},\ }\bibfield  {title} {\bibinfo {title}
  {Multimode storage and retrieval of microwave fields in a spin ensemble},\
  }\href {https://doi.org/10.1103/PhysRevX.4.021049} {\bibfield  {journal}
  {\bibinfo  {journal} {Phys. Rev. X}\ }\textbf {\bibinfo {volume} {4}},\
  \bibinfo {pages} {021049} (\bibinfo {year} {2014})}\BibitemShut {NoStop}%
\bibitem [{\citenamefont {Grezes}\ \emph {et~al.}(2015)\citenamefont {Grezes},
  \citenamefont {Julsgaard}, \citenamefont {Kubo}, \citenamefont {Ma},
  \citenamefont {Stern}, \citenamefont {Bienfait}, \citenamefont {Nakamura},
  \citenamefont {Isoya}, \citenamefont {Onoda}, \citenamefont {Ohshima},
  \citenamefont {Jacques}, \citenamefont {Vion}, \citenamefont {Esteve},
  \citenamefont {Liu}, \citenamefont {M\o{}lmer},\ and\ \citenamefont
  {Bertet}}]{Grezes2015}%
  \BibitemOpen
  \bibfield  {author} {\bibinfo {author} {\bibfnamefont {C.}~\bibnamefont
  {Grezes}}, \bibinfo {author} {\bibfnamefont {B.}~\bibnamefont {Julsgaard}},
  \bibinfo {author} {\bibfnamefont {Y.}~\bibnamefont {Kubo}}, \bibinfo {author}
  {\bibfnamefont {W.~L.}\ \bibnamefont {Ma}}, \bibinfo {author} {\bibfnamefont
  {M.}~\bibnamefont {Stern}}, \bibinfo {author} {\bibfnamefont
  {A.}~\bibnamefont {Bienfait}}, \bibinfo {author} {\bibfnamefont
  {K.}~\bibnamefont {Nakamura}}, \bibinfo {author} {\bibfnamefont
  {J.}~\bibnamefont {Isoya}}, \bibinfo {author} {\bibfnamefont
  {S.}~\bibnamefont {Onoda}}, \bibinfo {author} {\bibfnamefont
  {T.}~\bibnamefont {Ohshima}}, \bibinfo {author} {\bibfnamefont
  {V.}~\bibnamefont {Jacques}}, \bibinfo {author} {\bibfnamefont
  {D.}~\bibnamefont {Vion}}, \bibinfo {author} {\bibfnamefont {D.}~\bibnamefont
  {Esteve}}, \bibinfo {author} {\bibfnamefont {R.~B.}\ \bibnamefont {Liu}},
  \bibinfo {author} {\bibfnamefont {K.}~\bibnamefont {M\o{}lmer}},\ and\
  \bibinfo {author} {\bibfnamefont {P.}~\bibnamefont {Bertet}},\ }\bibfield
  {title} {\bibinfo {title} {Storage and retrieval of microwave fields at the
  single-photon level in a spin ensemble},\ }\href
  {https://doi.org/10.1103/PhysRevA.92.020301} {\bibfield  {journal} {\bibinfo
  {journal} {Phys. Rev. A}\ }\textbf {\bibinfo {volume} {92}},\ \bibinfo
  {pages} {020301} (\bibinfo {year} {2015})}\BibitemShut {NoStop}%
\bibitem [{\citenamefont {Flurin}\ \emph {et~al.}(2015)\citenamefont {Flurin},
  \citenamefont {Roch}, \citenamefont {Pillet}, \citenamefont {Mallet},\ and\
  \citenamefont {Huard}}]{Mallet2015}%
  \BibitemOpen
  \bibfield  {author} {\bibinfo {author} {\bibfnamefont {E.}~\bibnamefont
  {Flurin}}, \bibinfo {author} {\bibfnamefont {N.}~\bibnamefont {Roch}},
  \bibinfo {author} {\bibfnamefont {J.~D.}\ \bibnamefont {Pillet}}, \bibinfo
  {author} {\bibfnamefont {F.}~\bibnamefont {Mallet}},\ and\ \bibinfo {author}
  {\bibfnamefont {B.}~\bibnamefont {Huard}},\ }\bibfield  {title} {\bibinfo
  {title} {Superconducting quantum node for entanglement and storage of
  microwave radiation},\ }\href
  {https://doi.org/10.1103/PhysRevLett.114.090503} {\bibfield  {journal}
  {\bibinfo  {journal} {Phys. Rev. Lett.}\ }\textbf {\bibinfo {volume} {114}},\
  \bibinfo {pages} {090503} (\bibinfo {year} {2015})}\BibitemShut {NoStop}%
\bibitem [{\citenamefont {Reagor}\ \emph {et~al.}(2016)\citenamefont {Reagor},
  \citenamefont {Pfaff}, \citenamefont {Axline}, \citenamefont {Heeres},
  \citenamefont {Ofek}, \citenamefont {Sliwa}, \citenamefont {Holland},
  \citenamefont {Wang}, \citenamefont {Blumoff}, \citenamefont {Chou},
  \citenamefont {Hatridge}, \citenamefont {Frunzio}, \citenamefont {Devoret},
  \citenamefont {Jiang},\ and\ \citenamefont {Schoelkopf}}]{Schoelkopf2016}%
  \BibitemOpen
  \bibfield  {author} {\bibinfo {author} {\bibfnamefont {M.}~\bibnamefont
  {Reagor}}, \bibinfo {author} {\bibfnamefont {W.}~\bibnamefont {Pfaff}},
  \bibinfo {author} {\bibfnamefont {C.}~\bibnamefont {Axline}}, \bibinfo
  {author} {\bibfnamefont {R.~W.}\ \bibnamefont {Heeres}}, \bibinfo {author}
  {\bibfnamefont {N.}~\bibnamefont {Ofek}}, \bibinfo {author} {\bibfnamefont
  {K.}~\bibnamefont {Sliwa}}, \bibinfo {author} {\bibfnamefont
  {E.}~\bibnamefont {Holland}}, \bibinfo {author} {\bibfnamefont
  {C.}~\bibnamefont {Wang}}, \bibinfo {author} {\bibfnamefont {J.}~\bibnamefont
  {Blumoff}}, \bibinfo {author} {\bibfnamefont {K.}~\bibnamefont {Chou}},
  \bibinfo {author} {\bibfnamefont {M.~J.}\ \bibnamefont {Hatridge}}, \bibinfo
  {author} {\bibfnamefont {L.}~\bibnamefont {Frunzio}}, \bibinfo {author}
  {\bibfnamefont {M.~H.}\ \bibnamefont {Devoret}}, \bibinfo {author}
  {\bibfnamefont {L.}~\bibnamefont {Jiang}},\ and\ \bibinfo {author}
  {\bibfnamefont {R.~J.}\ \bibnamefont {Schoelkopf}},\ }\bibfield  {title}
  {\bibinfo {title} {Quantum memory with millisecond coherence in circuit
  qed},\ }\href {https://doi.org/10.1103/PhysRevB.94.014506} {\bibfield
  {journal} {\bibinfo  {journal} {Phys. Rev. B}\ }\textbf {\bibinfo {volume}
  {94}},\ \bibinfo {pages} {014506} (\bibinfo {year} {2016})}\BibitemShut
  {NoStop}%
\bibitem [{\citenamefont {Yin}\ \emph {et~al.}(2013)\citenamefont {Yin},
  \citenamefont {Chen}, \citenamefont {Sank}, \citenamefont {O'Malley},
  \citenamefont {White}, \citenamefont {Barends}, \citenamefont {Kelly},
  \citenamefont {Lucero}, \citenamefont {Mariantoni}, \citenamefont {Megrant},
  \citenamefont {Neill}, \citenamefont {Vainsencher}, \citenamefont {Wenner},
  \citenamefont {Korotkov}, \citenamefont {Cleland},\ and\ \citenamefont
  {Martinis}}]{Martinis2013}%
  \BibitemOpen
  \bibfield  {author} {\bibinfo {author} {\bibfnamefont {Y.}~\bibnamefont
  {Yin}}, \bibinfo {author} {\bibfnamefont {Y.}~\bibnamefont {Chen}}, \bibinfo
  {author} {\bibfnamefont {D.}~\bibnamefont {Sank}}, \bibinfo {author}
  {\bibfnamefont {P.~J.~J.}\ \bibnamefont {O'Malley}}, \bibinfo {author}
  {\bibfnamefont {T.~C.}\ \bibnamefont {White}}, \bibinfo {author}
  {\bibfnamefont {R.}~\bibnamefont {Barends}}, \bibinfo {author} {\bibfnamefont
  {J.}~\bibnamefont {Kelly}}, \bibinfo {author} {\bibfnamefont
  {E.}~\bibnamefont {Lucero}}, \bibinfo {author} {\bibfnamefont
  {M.}~\bibnamefont {Mariantoni}}, \bibinfo {author} {\bibfnamefont
  {A.}~\bibnamefont {Megrant}}, \bibinfo {author} {\bibfnamefont
  {C.}~\bibnamefont {Neill}}, \bibinfo {author} {\bibfnamefont
  {A.}~\bibnamefont {Vainsencher}}, \bibinfo {author} {\bibfnamefont
  {J.}~\bibnamefont {Wenner}}, \bibinfo {author} {\bibfnamefont {A.~N.}\
  \bibnamefont {Korotkov}}, \bibinfo {author} {\bibfnamefont {A.~N.}\
  \bibnamefont {Cleland}},\ and\ \bibinfo {author} {\bibfnamefont {J.~M.}\
  \bibnamefont {Martinis}},\ }\bibfield  {title} {\bibinfo {title} {Catch and
  release of microwave photon states},\ }\href
  {https://doi.org/10.1103/PhysRevLett.110.107001} {\bibfield  {journal}
  {\bibinfo  {journal} {Phys. Rev. Lett.}\ }\textbf {\bibinfo {volume} {110}},\
  \bibinfo {pages} {107001} (\bibinfo {year} {2013})}\BibitemShut {NoStop}%
\bibitem [{\citenamefont {Wenner}\ \emph {et~al.}(2014)\citenamefont {Wenner},
  \citenamefont {Yin}, \citenamefont {Chen}, \citenamefont {Barends},
  \citenamefont {Chiaro}, \citenamefont {Jeffrey}, \citenamefont {Kelly},
  \citenamefont {Megrant}, \citenamefont {Mutus}, \citenamefont {Neill},
  \citenamefont {O'Malley}, \citenamefont {Roushan}, \citenamefont {Sank},
  \citenamefont {Vainsencher}, \citenamefont {White}, \citenamefont {Korotkov},
  \citenamefont {Cleland},\ and\ \citenamefont {Martinis}}]{Martinis2014}%
  \BibitemOpen
  \bibfield  {author} {\bibinfo {author} {\bibfnamefont {J.}~\bibnamefont
  {Wenner}}, \bibinfo {author} {\bibfnamefont {Y.}~\bibnamefont {Yin}},
  \bibinfo {author} {\bibfnamefont {Y.}~\bibnamefont {Chen}}, \bibinfo {author}
  {\bibfnamefont {R.}~\bibnamefont {Barends}}, \bibinfo {author} {\bibfnamefont
  {B.}~\bibnamefont {Chiaro}}, \bibinfo {author} {\bibfnamefont
  {E.}~\bibnamefont {Jeffrey}}, \bibinfo {author} {\bibfnamefont
  {J.}~\bibnamefont {Kelly}}, \bibinfo {author} {\bibfnamefont
  {A.}~\bibnamefont {Megrant}}, \bibinfo {author} {\bibfnamefont {J.~Y.}\
  \bibnamefont {Mutus}}, \bibinfo {author} {\bibfnamefont {C.}~\bibnamefont
  {Neill}}, \bibinfo {author} {\bibfnamefont {P.~J.~J.}\ \bibnamefont
  {O'Malley}}, \bibinfo {author} {\bibfnamefont {P.}~\bibnamefont {Roushan}},
  \bibinfo {author} {\bibfnamefont {D.}~\bibnamefont {Sank}}, \bibinfo {author}
  {\bibfnamefont {A.}~\bibnamefont {Vainsencher}}, \bibinfo {author}
  {\bibfnamefont {T.~C.}\ \bibnamefont {White}}, \bibinfo {author}
  {\bibfnamefont {A.~N.}\ \bibnamefont {Korotkov}}, \bibinfo {author}
  {\bibfnamefont {A.~N.}\ \bibnamefont {Cleland}},\ and\ \bibinfo {author}
  {\bibfnamefont {J.~M.}\ \bibnamefont {Martinis}},\ }\bibfield  {title}
  {\bibinfo {title} {Catching time-reversed microwave coherent state photons
  with 99.4$\%$ absorption efficiency},\ }\href@noop {} {\bibfield  {journal}
  {\bibinfo  {journal} {Phys. Rev. Lett.}\ }\textbf {\bibinfo {volume} {112}},\
  \bibinfo {pages} {210501} (\bibinfo {year} {2014})}\BibitemShut {NoStop}%
\bibitem [{\citenamefont {Mirrahimi}\ \emph {et~al.}(2014)\citenamefont
  {Mirrahimi}, \citenamefont {Leghtas}, \citenamefont {Albert}, \citenamefont
  {Touzard}, \citenamefont {Schoelkopf}, \citenamefont {Jiang},\ and\
  \citenamefont {Devoret}}]{Mirrahimi2014}%
  \BibitemOpen
  \bibfield  {author} {\bibinfo {author} {\bibfnamefont {M.}~\bibnamefont
  {Mirrahimi}}, \bibinfo {author} {\bibfnamefont {Z.}~\bibnamefont {Leghtas}},
  \bibinfo {author} {\bibfnamefont {V.~V.}\ \bibnamefont {Albert}}, \bibinfo
  {author} {\bibfnamefont {S.}~\bibnamefont {Touzard}}, \bibinfo {author}
  {\bibfnamefont {R.~J.}\ \bibnamefont {Schoelkopf}}, \bibinfo {author}
  {\bibfnamefont {L.}~\bibnamefont {Jiang}},\ and\ \bibinfo {author}
  {\bibfnamefont {M.~H.}\ \bibnamefont {Devoret}},\ }\bibfield  {title}
  {\bibinfo {title} {Dynamically protected cat-qubits: a new paradigm for
  universal quantum computation},\ }\href@noop {} {\bibfield  {journal}
  {\bibinfo  {journal} {New Journal of Physics}\ }\textbf {\bibinfo {volume}
  {16}},\ \bibinfo {pages} {045014} (\bibinfo {year} {2014})}\BibitemShut
  {NoStop}%
\bibitem [{\citenamefont {Duivenvoorden}\ \emph {et~al.}(2017)\citenamefont
  {Duivenvoorden}, \citenamefont {Terhal},\ and\ \citenamefont
  {Weigand}}]{Weigand2017}%
  \BibitemOpen
  \bibfield  {author} {\bibinfo {author} {\bibfnamefont {K.}~\bibnamefont
  {Duivenvoorden}}, \bibinfo {author} {\bibfnamefont {B.~M.}\ \bibnamefont
  {Terhal}},\ and\ \bibinfo {author} {\bibfnamefont {D.}~\bibnamefont
  {Weigand}},\ }\bibfield  {title} {\bibinfo {title} {Single-mode displacement
  sensor},\ }\href {https://doi.org/10.1103/PhysRevA.95.012305} {\bibfield
  {journal} {\bibinfo  {journal} {Phys. Rev. A}\ }\textbf {\bibinfo {volume}
  {95}},\ \bibinfo {pages} {012305} (\bibinfo {year} {2017})}\BibitemShut
  {NoStop}%
\bibitem [{\citenamefont {Moiseev}\ \emph {et~al.}(2017)\citenamefont
  {Moiseev}, \citenamefont {Gubaidullin}, \citenamefont {Kirillov},
  \citenamefont {Latypov}, \citenamefont {Perminov}, \citenamefont
  {Petrovnin},\ and\ \citenamefont {Sherstyukov}}]{Moiseev2017}%
  \BibitemOpen
  \bibfield  {author} {\bibinfo {author} {\bibfnamefont {S.~A.}\ \bibnamefont
  {Moiseev}}, \bibinfo {author} {\bibfnamefont {F.~F.}\ \bibnamefont
  {Gubaidullin}}, \bibinfo {author} {\bibfnamefont {R.~S.}\ \bibnamefont
  {Kirillov}}, \bibinfo {author} {\bibfnamefont {R.~R.}\ \bibnamefont
  {Latypov}}, \bibinfo {author} {\bibfnamefont {N.~S.}\ \bibnamefont
  {Perminov}}, \bibinfo {author} {\bibfnamefont {K.~V.}\ \bibnamefont
  {Petrovnin}},\ and\ \bibinfo {author} {\bibfnamefont {O.~N.}\ \bibnamefont
  {Sherstyukov}},\ }\bibfield  {title} {\bibinfo {title} {Multiresonator
  quantum memory},\ }\href@noop {} {\bibfield  {journal} {\bibinfo  {journal}
  {Phys. Rev. A}\ }\textbf {\bibinfo {volume} {95}},\ \bibinfo {pages} {012338}
  (\bibinfo {year} {2017})}\BibitemShut {NoStop}%
\bibitem [{\citenamefont {Afzelius}\ \emph {et~al.}(2009)\citenamefont
  {Afzelius}, \citenamefont {Simon}, \citenamefont {de~Riedmatten},\ and\
  \citenamefont {Gisin}}]{Afzelius2009}%
  \BibitemOpen
  \bibfield  {author} {\bibinfo {author} {\bibfnamefont {M.}~\bibnamefont
  {Afzelius}}, \bibinfo {author} {\bibfnamefont {C.}~\bibnamefont {Simon}},
  \bibinfo {author} {\bibfnamefont {H.}~\bibnamefont {de~Riedmatten}},\ and\
  \bibinfo {author} {\bibfnamefont {N.}~\bibnamefont {Gisin}},\ }\bibfield
  {title} {\bibinfo {title} {Multimode quantum memory based on atomic frequency
  combs},\ }\href {https://doi.org/10.1103/PhysRevA.79.052329} {\bibfield
  {journal} {\bibinfo  {journal} {Phys. Rev. A}\ }\textbf {\bibinfo {volume}
  {79}},\ \bibinfo {pages} {052329} (\bibinfo {year} {2009})}\BibitemShut
  {NoStop}%
\bibitem [{\citenamefont {Moiseev}\ \emph {et~al.}(2018)\citenamefont
  {Moiseev}, \citenamefont {Gerasimov}, \citenamefont {Latypov}, \citenamefont
  {Perminov}, \citenamefont {Petrovnin},\ and\ \citenamefont
  {Sherstyukov}}]{Moiseev2018}%
  \BibitemOpen
  \bibfield  {author} {\bibinfo {author} {\bibfnamefont {S.~A.}\ \bibnamefont
  {Moiseev}}, \bibinfo {author} {\bibfnamefont {K.~I.}\ \bibnamefont
  {Gerasimov}}, \bibinfo {author} {\bibfnamefont {R.~R.}\ \bibnamefont
  {Latypov}}, \bibinfo {author} {\bibfnamefont {N.~S.}\ \bibnamefont
  {Perminov}}, \bibinfo {author} {\bibfnamefont {K.~V.}\ \bibnamefont
  {Petrovnin}},\ and\ \bibinfo {author} {\bibfnamefont {O.~N.}\ \bibnamefont
  {Sherstyukov}},\ }\bibfield  {title} {\bibinfo {title} {Broadband
  multiresonator quantum memory-interface},\ }\href
  {https://doi.org/10.1038/s41598-018-21941-6} {\bibfield  {journal} {\bibinfo
  {journal} {Scientific Reports}\ }\textbf {\bibinfo {volume} {8}},\ \bibinfo
  {pages} {3982} (\bibinfo {year} {2018})}\BibitemShut {NoStop}%
\bibitem [{\citenamefont {Sandberg}\ \emph {et~al.}(2008)\citenamefont
  {Sandberg}, \citenamefont {Wilson}, \citenamefont {Persson}, \citenamefont
  {Bauch}, \citenamefont {Johansson}, \citenamefont {Shumeiko}, \citenamefont
  {Duty},\ and\ \citenamefont {Delsing}}]{Delsing2008}%
  \BibitemOpen
  \bibfield  {author} {\bibinfo {author} {\bibfnamefont {M.}~\bibnamefont
  {Sandberg}}, \bibinfo {author} {\bibfnamefont {C.~M.}\ \bibnamefont
  {Wilson}}, \bibinfo {author} {\bibfnamefont {F.}~\bibnamefont {Persson}},
  \bibinfo {author} {\bibfnamefont {T.}~\bibnamefont {Bauch}}, \bibinfo
  {author} {\bibfnamefont {G.}~\bibnamefont {Johansson}}, \bibinfo {author}
  {\bibfnamefont {V.}~\bibnamefont {Shumeiko}}, \bibinfo {author}
  {\bibfnamefont {T.}~\bibnamefont {Duty}},\ and\ \bibinfo {author}
  {\bibfnamefont {P.}~\bibnamefont {Delsing}},\ }\bibfield  {title} {\bibinfo
  {title} {Tuning the field in a microwave resonator faster than the photon
  lifetime},\ }\href {https://doi.org/10.1063/1.2929367} {\bibfield  {journal}
  {\bibinfo  {journal} {Applied Physics Letters}\ }\textbf {\bibinfo {volume}
  {92}},\ \bibinfo {pages} {203501} (\bibinfo {year} {2008})}\BibitemShut
  {NoStop}%
\bibitem [{sup()}]{supp}%
  \BibitemOpen
  \href@noop {} {\ \ \bibinfo {pages} {A detailed discussion can be found in
  the Supplementary Information.}}\BibitemShut {Stop}%
\bibitem [{\citenamefont {Kim}\ \emph {et~al.}(2019)\citenamefont {Kim},
  \citenamefont {Shrekenhamer}, \citenamefont {McElroy}, \citenamefont
  {Strikwerda},\ and\ \citenamefont {Alldredge}}]{Kim2019}%
  \BibitemOpen
  \bibfield  {author} {\bibinfo {author} {\bibfnamefont {S.}~\bibnamefont
  {Kim}}, \bibinfo {author} {\bibfnamefont {D.}~\bibnamefont {Shrekenhamer}},
  \bibinfo {author} {\bibfnamefont {K.}~\bibnamefont {McElroy}}, \bibinfo
  {author} {\bibfnamefont {A.}~\bibnamefont {Strikwerda}},\ and\ \bibinfo
  {author} {\bibfnamefont {J.}~\bibnamefont {Alldredge}},\ }\bibfield  {title}
  {\bibinfo {title} {Tunable superconducting cavity using superconducting
  quantum interference device metamaterials},\ }\href
  {https://doi.org/10.1038/s41598-019-40891-1} {\bibfield  {journal} {\bibinfo
  {journal} {Scientific Reports}\ }\textbf {\bibinfo {volume} {9}},\ \bibinfo
  {pages} {4630} (\bibinfo {year} {2019})}\BibitemShut {NoStop}%
\bibitem [{\citenamefont {de~Riedmatten}\ \emph {et~al.}(2008)\citenamefont
  {de~Riedmatten}, \citenamefont {Afzelius}, \citenamefont {Staudt},
  \citenamefont {Simon},\ and\ \citenamefont {Gisin}}]{Gisin2008}%
  \BibitemOpen
  \bibfield  {author} {\bibinfo {author} {\bibfnamefont {H.}~\bibnamefont
  {de~Riedmatten}}, \bibinfo {author} {\bibfnamefont {M.}~\bibnamefont
  {Afzelius}}, \bibinfo {author} {\bibfnamefont {M.~U.}\ \bibnamefont
  {Staudt}}, \bibinfo {author} {\bibfnamefont {C.}~\bibnamefont {Simon}},\ and\
  \bibinfo {author} {\bibfnamefont {N.}~\bibnamefont {Gisin}},\ }\bibfield
  {title} {\bibinfo {title} {A solid-state light–matter interface at the
  single-photon level},\ }\href@noop {} {\bibfield  {journal} {\bibinfo
  {journal} {Nature}\ }\textbf {\bibinfo {volume} {456}},\ \bibinfo {pages}
  {773} (\bibinfo {year} {2008})}\BibitemShut {NoStop}%
\bibitem [{\citenamefont {G\"undo\ifmmode~\breve{g}\else \u{g}\fi{}an}\ \emph
  {et~al.}(2015)\citenamefont {G\"undo\ifmmode~\breve{g}\else \u{g}\fi{}an},
  \citenamefont {Ledingham}, \citenamefont {Kutluer}, \citenamefont {Mazzera},\
  and\ \citenamefont {de~Riedmatten}}]{Hugues2015}%
  \BibitemOpen
  \bibfield  {author} {\bibinfo {author} {\bibfnamefont {M.}~\bibnamefont
  {G\"undo\ifmmode~\breve{g}\else \u{g}\fi{}an}}, \bibinfo {author}
  {\bibfnamefont {P.~M.}\ \bibnamefont {Ledingham}}, \bibinfo {author}
  {\bibfnamefont {K.}~\bibnamefont {Kutluer}}, \bibinfo {author} {\bibfnamefont
  {M.}~\bibnamefont {Mazzera}},\ and\ \bibinfo {author} {\bibfnamefont
  {H.}~\bibnamefont {de~Riedmatten}},\ }\bibfield  {title} {\bibinfo {title}
  {Solid state spin-wave quantum memory for time-bin qubits},\ }\href@noop {}
  {\bibfield  {journal} {\bibinfo  {journal} {Phys. Rev. Lett.}\ }\textbf
  {\bibinfo {volume} {114}},\ \bibinfo {pages} {230501} (\bibinfo {year}
  {2015})}\BibitemShut {NoStop}%
\end{thebibliography}

%

\end{document}